\documentclass[aps,prd,twocolumn,superscriptaddress]{revtex4-2}

\usepackage[colorlinks=true,citecolor=red,urlcolor=blue,linkcolor=blue]{hyperref}

\usepackage{graphicx}
\usepackage{dcolumn}
\usepackage{bm}
\usepackage{color}

\begin{document}

\preprint{APS/123-QED}

\title{
Exploring Time Delay Interferometry Ranging  as a Practical Ranging Approach in the Bayesian Framework
}

\author{Minghui Du}
\email{duminghui@imech.ac.cn}
\affiliation{Institute of Mechanics, Chinese Academy of Sciences, Beĳing 100190, China.}
\author{Pengzhan Wu}
\affiliation{Institute of Mechanics, Chinese Academy of Sciences, Beĳing 100190, China.}
\author{Ziren Luo}
\affiliation{Institute of Mechanics, Chinese Academy of Sciences, Beĳing 100190, China.}
\affiliation{Hangzhou Institute for Advanced Study, University of Chinese Academy of Sciences, Hangzhou 310124, China.}
\author{Peng Xu}
\email{xupeng@imech.ac.cn}
\affiliation{Institute of Mechanics, Chinese Academy of Sciences, Beĳing 100190, China.}
\affiliation{Lanzhou Center of Theoretical Physics, Lanzhou University, Lanzhou 730000, China.}
\affiliation{Hangzhou Institute for Advanced Study, University of Chinese Academy of Sciences, Hangzhou 310124, China.}




\date{\today}

\begin{abstract}
Time Delay Interferometry (TDI) is an indispensable step in the whole data processing procedure of space-based gravitational wave detection, as it mitigates the overwhelming laser frequency noise, which would otherwise completely bury the gravitational wave signals. Knowledge on the inter-spacecraft optical paths (i.e. delays) is one of the key elements of TDI. Conventional method for inter-spacecraft ranging mainly relies on the pseudo-random noise (PRN) code signal modulated onto the lasers. To ensure the reliability and robustness of this ranging information, it would be highly beneficial to develop other methods which could serve as cross-validations or backups. This paper explores the practical implementation of an alternative data-driven approach - time delay interferometry ranging (TDIR) - as a ranging technique independent of the PRN signal. Distinguished from previous research, our TDIR algorithm significantly relaxes the stringent requirement for clock synchronization imposed by traditional TDI procedure. By framing TDIR as a Bayesian parameter estimation problem and employing a general polynomial parametrization, we demonstrate our algorithm with simulated data based on the numerical orbit of Taiji. In the presence of laser frequency noise and secondary noises, the estimated median values of delays are only 5.28 ns away from the ground truths, capable of suppressing laser frequency noise to the desired level. Additionally, we have also analysed the requirements of mitigating optical bench noise and clock noise on TDIR, and presented an illustrative example for the impact of laser locking.
\end{abstract}

\maketitle


\section{Introduction}
\label{sec:introduction}
The first detection of gravitational wave (GW)~\cite{PhysRevLett.116.061102} marked the beginning of the gravitational-wave astronomy era. 
The LIGO-VIRGO-KAGRA ground-based GW detector network, whose target frequencies ranges from  1 Hz to 1 kHz, has detected over a hundred GW events originating from compact binaries.
In contrast, in the upcoming decade, space-based detectors such as the Laser Interferometer Space Antenna (LISA)~\cite{Barausse:2020rsu}, Taiji~\cite{Taiji,Taiji1,Taiji2}, and Tianqin~\cite{TianQin:2015yph}, aim to detect GWs in the 0.1 mHz - 1 Hz frequency band associated with enormous astrophysical and cosmological procedures~\cite{TaijiSO}, such as the mergers of massive black hole binaries~\cite{SO_MBHB}, the orbiting  of compact binaries  within the Milky Way~\cite{SO_GB}, the extreme mass ratio inspirals, and the first-order phase transition in the early universe~\cite{SO_FOPT}, etc. As a representative example, Taiji consists of three  spacecrafts (S/Cs) connected by bidirectional laser links, with a baseline armlength of 3 million kilometers. The whole constellation   orbits around the sun and leads the Earth by about 20 degrees.

Albeit overcoming the limitations of the Earth's scale and seismic noise, space-based GW detectors subject to  laser frequency noise (LFN) which is coupled with 
the inequality of armlengths. LFN in the raw interferometric measurements can be 6 - 8 orders stronger than  the targeted GW signals. 
To suppress this overwhelming noise source, the TDI technique was proposed to synthesise a virtual equal-arm interferometer in the post-processing stage~\cite{1stTDI,1stTDIAET,1stTDIWTN}. Generally, the construction of TDI combinations involves appropriately delaying and combining the readouts of  interferometors, including the science (i.e. inter-spacecraft), reference and test-mass interferometors. According to the capability of LFN suppression, TDI combinations can be  categorized into the first and second generations. For an ideal stationary unequal-arm constellation, perfect cancellation of LFN can be achieved by the first-generation combinations. While 
for LISA and Taiji whose armlengths vary at rates  up to several m/s, it is necessary to employ the second-generation TDI combinations to account for both the inequality and time dependence of the  armlengths~\cite{TDI2}.

In addition to the interferometric measurements (also known as the beatnote signals), the inter-spacecraft optical paths (i.e. delays)  constitute another crucial input for TDI. 
Technically, the inter-spacecraft ranging  can be done by modulating  the  PRN code onto the lasers using approximately $1 \%$ of the  power~\cite{Esteban_2009,Esteban_2010,Sutton:10,Heinzel_2011,TaijiPRN}. The resultant PRN range (PRNR) may suffer from several imperfections, such as ranging noise, ambiguity, modulation noise, and on-board delays, etc.
TDI places stringent requirement on the precision of delays
($\mathcal{O}(10) \sim \mathcal{O}(100) \ {\rm ns}$, depending on the TDI channels~\cite{TDI_ranging_precision,TintoLivingReview2020}), 
so the practical application of PRNR necessitates careful processing through the combination  with other ranging observables~\cite{ranging_fusion}, based on  a thorough investigation and comprehensive understanding of the technical details. 
Due to these technical imperfections and complexities, as well as  the fundamental importance of TDI for space-based GW detection, the development of other ranging methods that are independent of PRN signal and its processing step is also desirable. Such method could serve as a risk mitigation or cross verification, and hence improve the reliability and robustness of  ranging.

TDIR is a data-driven approach of inter-spacecraft ranging, which does not directly measure the delays, but instead searches for the delays that  minimize the residual LFN in the TDI combinations~\cite{TDIR_Tinto}. 
Ref.~\cite{TDIR_Tinto} proposed the basic concept of TDIR, construing it as an optimization problem aimed at minimizing the integrated  power of a single TDI channel. 
When the  delays used in TDI are close to their true values, LFN  can be suppressed well below the secondary noises (i.e. the optical measurement system (OMS) noises and residual acceleration noises of the test masses).
According to the principles of  Bayesian statistical inference, an optimal estimator can be formulated by whitening the power of TDI channels using the noise power spectral density (PSD) of secondary noises in the Fourier domain. 
Researches based on this idea view TDIR as a Bayesian parameter estimation problem, which not only yields the best estimate of  delays, but also outputs the confidence interval for the estimate, facilitating the assessment of ranging precision. 
Ref.~\cite{Bayesian_TDIR1} formulated the  likelihood of TDIR for he  first-generation Michelson channels $\{X, Y, Z\}$, ran a Markov Chain Monte Carlo (MCMC) search on the simulated data, and constrained the delays to a $\mathcal{O}(10) \ {\rm ns}$ precision.  
In a following up research~\cite{Bayesian_TDIR2}, the same algorithm was extended to the second-generation TDI and time-varying delays, presuming that the delays could be parameterized by  Keplerian orbital elements. No-bias estimate of the parameters was obtained on LFN-only simulations. Complementarily, a theoretical analysis on the precision of  TDIR was made by Ref.~\cite{Martin_PhD_thesis} using the Fisher matrix formalism. 
In addition to these theoretical work,  the verification of a simplified TDIR algorithm  using the data of GRACE Follow-On satellites was described in Ref.~\cite{TDIR_GFO}.

In the most recent research regarding Bayesian TDIR~\cite{Bayesian_TDIR2}, the secondary noises, which may drive the estimated parameters away from their true values, are omitted in the simulation. 
To further investigate the performance of TDIR when secondary noises are also present, 
in this paper, we apply TDIR  on simulated data  including the realizations of  LFN and secondary noises. 
For the sake of practicality,  the data are generated based on the  numerical orbit of Taiji. 
More importantly, since we aim to develop TDIR into a ranging technique independent of the PRN signal, the implementation of it cannot directly or indirectly rely  on the PRN signal. This implies that we need to make some modifications to the algorithm, especially considering its relationship with clock synchronization, as will be explained in the following.

A subtle aspect about TDIR is its relationship with  clock synchronization. 
The laser beatnotes and other measurements are recorded with respect to the clocks (such as ultra-stable oscillators, USOs) hosted by the three S/Cs, each exhibiting a unique deviation from the proper time due to instrumental imperfections including long-term deterministic drifts and random jitters. 
To relate the  data  with astrophysical events and in-orbit incidents, thereby enabling the estimation of GW source parameters and monitoring the  status of detector, it is essential to adjust the timestamps of data from the three S/Cs to a common time frame  (i.e. clock synchronization). 
In the context of TDI, ranging and synchronization are closely related for at least two reasons.
Firstly, during the construction of TDI channels, the time frame of beatnote signals must be consistent with the delays applied on them. For example, if the signals are synchronized to the Solar system barycentric coordinate time (TCB)~\cite{timescale,PhysRevD.69.082001,YanWang_PhD_thesis}, then the delays must be the light travel time (LTT)  defined as the difference of TCB times between photon reception  and emission. Ref.~\cite{TintoLivingReview2020,PhysRevD.70.081101} predicted that TDI requires clock synchronization to reach a $\mathcal{O}(10)$-$\mathcal{O}(100) \ {\rm ns}$ precision. 
While, if the signals are the unsynchronized raw measurements, then the delays should be accordingly the PRNR, defined as the difference of onboard clock readings between photon reception and emission, which encodes the information of both LTT and clock deviations (i.e. the ``TDI without synchronization'' scheme, see Ref.~\cite{TDINoSync} for more detailed  explanations). 
Secondly, in the conventional data processing pipeline~\cite{PhysRevD.90.064016}, both the nanosecond level high-precision ranging and clock synchronization rely on the PRN signal. An important aspect in the processing of   PRN signal  is to  disentangle the  LTTs and clock deviations using Kalman-like filters, based on  the modeling of clocks, orbital motions and relativistic corrections. 

Most of the previous studies~\cite{TDIR_Tinto,Bayesian_TDIR1,Bayesian_TDIR2} illustrated TDIR on ideally synchronized data (e.g. synchronization is explicitly stated in Ref.~\cite{TDIR_Tinto}, and Ref.~\cite{Bayesian_TDIR2} modeled the delays as LTTs, hence implicitly assuming synchronized to TCB). 
As is explained, these implementations of TDIR  inherently relied on PRNR. 
Therefore, using ideally synchronized data contradicts our intention to 
avoid the dependence of TDIR on the PRN signal.
Ref.~\cite{ranging_fusion} applied TDIR to unsynchronized raw measurements, while the authors  only treated it  as an auxiliary step in the so-called ``ranging sensor fusion'' processing, to correct the bias and ambiguity of PRNR.   
In this paper, we investigate the feasibility  of TDIR on data that has  been preliminarily synchronized to TCB with  a $0.1 \ {\rm ms}$  precision (still failing to meet  the requirement of TDI) using the ground tracking data.
As will be explained in Sec.~\ref{sec:demonstration}, in principle, TDIR  also applies to completely unsynchronized data. 
However, synchronizing to TCB (or other global time frames) is essential for the sake of GW source parameter estimation.
For the Taiji mission, the precision requirement on clock synchronization imposed by astrophysical data analysis is still  under on-going research guided by the  scientific objectives, but it is certainly much less stringent than the that imposed by LFN suppression.
Moreover, this preliminary synchronization also offers the opportunity of combining different TDI channels to achieve better  constraint on the delay parameters. 
Accordingly, to apply TDIR on unsynchronized data, we need to adopt  a more general parametrization (other than constant delays or Keplerian orbital elements).

Except for LFN, The complete TDI procedure is also known to mitigate the noises induced by the random displacements of optical benches (OBs) relative to the local inertial frames (i.e. OB noises), as well as   the noises originating from  the jitters of on board clocks (i.e. clock noises). Mitigation of the former noise is accomplished through the construction of test-mass-to-test-mass interferometry (intermediate variable $\xi$), and has been included in previous studies. The latter, however, which should be dealt with an extra clock noise reduction step subsequent to the combination of TDI channels using the information of clock sidebands~\cite{TintoClockNoise,clock_jitter_reduction}, has received scant attention in the literature about TDIR. 
In fact, the residual clock noise  in the second-generation TDI channel is still 2 - 3 orders larger than the tolerable noise level. If the clock noise reduction step is omitted, whitening the data using only the PSD of secondary noises would be incorrect. Therefore, we suggest that the practical implementation of TDIR should encompass this crucial step. Furthermore, through theoretical analysis and simulation, we investigate the requirements  of suppressing  OB and clock noises on the precision of TDIR.

In summary, this paper is dedicated to exploring TDIR as an independent  ranging technique, under the realistic scenarios including  LFN, secondary noises, OB noise, clock noise, orbital motion, and clock desynchronization. The whole TDI procedure based on our TDIR algorithm  is free of the imperfections of PRNR, as well as the sophisticated PRNR processing steps, and has more relaxed  requirement on clock synchronization.  The content of this paper is arranged as follows: 
Sec.~\ref{sec:demonstration}  explains theoretically the principle of applying   TDIR to unsynchronized data, and analyzes the couplings of  ranging errors with  LFN, OB noise and clock noise, thereby deriving the requirements of suppressing these noises on TDIR. 
Furthermore, the parametrization form of  delays is proposed and  justified based on the numerical orbit of Taiji. Sec.~\ref{sec:Bayesian} then  explains how to formulate TDIR as a Bayesian parameter estimation problem, and analytically estimates the  accuracy of parameters by means of Fisher matrix. The specific settings for data simulation and MCMC sampling are detailed in Sec.~\ref{sec:Simulation}, followed by the discussions on the results.   We draw our conclusion and outlook for future work in Sec.~\ref{sec:conclusions}. In addition, the models of instrumental noises and an illustrative example regarding the impact of laser locking are presented in Appendix~\ref{sec:appendix} and \ref{sec:locking}, respectively.

\section{demonstration of the principles}\label{sec:demonstration}
\subsection{TDIR on unsynchronized data}\label{subsec:principle}

In order to express the same  signal in  different time frames, 
for an arbitrary time series $f(\tau)$ (e.g. phase signal, clock reading, delay, etc.)
we follow the convention introduced by Ref.~\cite{LISA_instrument} to  add a superscript to $f$ to denote the time frame of argument $\tau$. 
For example, $f^t(\tau)$ is expressed in TCB, while $f^{\hat \tau_i}(\tau)$ is timestamped by the clock of  S/C$_i$. 
TCB is a global time frame for events in the solar system,  and the clock times are  the only times directly measurable by the clocks (such as USOs) onborad the S/Cs. 
When describing  the same event,  the latter  differs from the former due to relativistic effects induced by the gravitational field and orbital motions of S/Cs~\cite{timescale} 
, as well as instrumental imperfections, including  long-term, deterministic clock drifts and random  jitters. 
The relationship between clock time and TCB can be expressed as
\begin{equation}
    \hat{\tau}_i^t(\tau) = \tau + \delta \hat{\tau}_i^t(\tau),
\end{equation}
where the total deviation with respect to  TCB is grounded in one term $\delta \hat{\tau}_i^t(\tau)$. The conversion of $f$ between different time frames thus follows 
{\color{black}\begin{equation}\label{eq:time_frame_conversion}
    f^{\hat{\tau}_i}(\tau) = f^t\left[t^{\hat{\tau}_i}(\tau)\right],
\end{equation}}
 
On the other hand, the ``delays'' that appear in the expressions of beatnote signals and  TDI combinations  are defined  as the time difference between  photon reception and emission, recorded in a certain time frame. 
The time frame corresponding to LTT (denoted as $d^t_{ij}(\tau)$) is  TCB. As laser propagates from S/C$_j$ to S/C$_i$, we have
\begin{equation}
    d_{ij}^t(\tau) = t_{i, {\rm reception}} - t_{j, {\rm emission}},
\end{equation}
with $t_{i, {\rm reception}} = \tau$, meaning that the photon arrives at S/C$_i$ when the TCB time is $\tau$.
Throughout this paper we follow the prescription of Ref.~\cite{relativistic_link} to calculate LTT to the 1st post-Newtonian order. 
Meanwhile, the PRNR (dubbed $\hat{d}_{ij}^{\hat{tau}_i}(\tau)$, neglecting the technical imperfections such as ranging noise, ambiguity and bias, precisely tracks the onboard clock difference between  laser reception and emission:
\begin{eqnarray}
    \hat{d}_{ij}^t(\tau) &=& \hat{\tau}_i^t(\tau) - \hat{\tau}_j^t[\tau - d^t_{ij}(\tau)] \label{eq:mpr_t}\nonumber \\
    &=& d^t_{ij}(\tau) + \delta \hat{\tau}_i^t(\tau) - \delta \hat{\tau}_j^t\left[\tau - d^t_{ij}(\tau)\right], \\
    \hat{d}^{\hat{\tau}_i}_{ij}(\tau) &=& \hat{d}^t_{ij}\left[t^{\hat{\tau}_i}(\tau)\right], \label{eq:mpr_tau}
\end{eqnarray} 
which incorporates the information of both LTT and differential clock deviations between S/Cs. The first equation is expressed in TCB, and the second gives its transformation to the clock time of receiving S/C, as an application of Eq.~(\ref{eq:time_frame_conversion}). 
To effectively suppress LFN with TDI on unsynchronized data, similar to the ``TDI without clock synchronization'' scheme~\cite{TDINoSync}, 
the objective of TDIR can be interpreted as searching for the functional form that fits $\hat{d}_{ij}^{\hat{\tau}_i}(\tau)$ best. However, perfect reconstruction of $\hat{d}_{ij}^{\hat{\tau}_i}(\tau)$ with a fitting function is not practical  due to the presence of random clock jitters in $\delta \hat{\tau}_i^t$.  
 Functional forms with limited number of parameters (as few as possible to ensure the efficiency of MCMC sampling) can at most fit the slowly-varying, deterministic part of $\hat{d}_{ij}^{\hat{\tau}_i}(\tau)$.    
We refer to this variable as the ``fiducial'' TDIR $\bar{d}_{ij}^{\hat{\tau}_i}(\tau)$. To be distinguished,   the ``parametrized'' TDIR that will be fitted using the interferometric data is dubbed $R_{ij}(\tau, \bm{\theta})$. The fiducial TDIR will be treated as the ``ground truth'' of our searching  algorithm, and can be calculated via Eq.~(\ref{eq:mpr_tau}) assuming that there is no random clock jitters. Meanwhile, the parametrized TDIR contains unknown model parameters to be fitted using the beatnote signals.

Next, we explain the principle of suppressing LFN in unsynchronized data using the aforementioned TDIR, starting with the equations of interferometric measurements  expressed in terms of TDIR.

\begin{figure}
    \centering
    \includegraphics[width=0.5\textwidth]{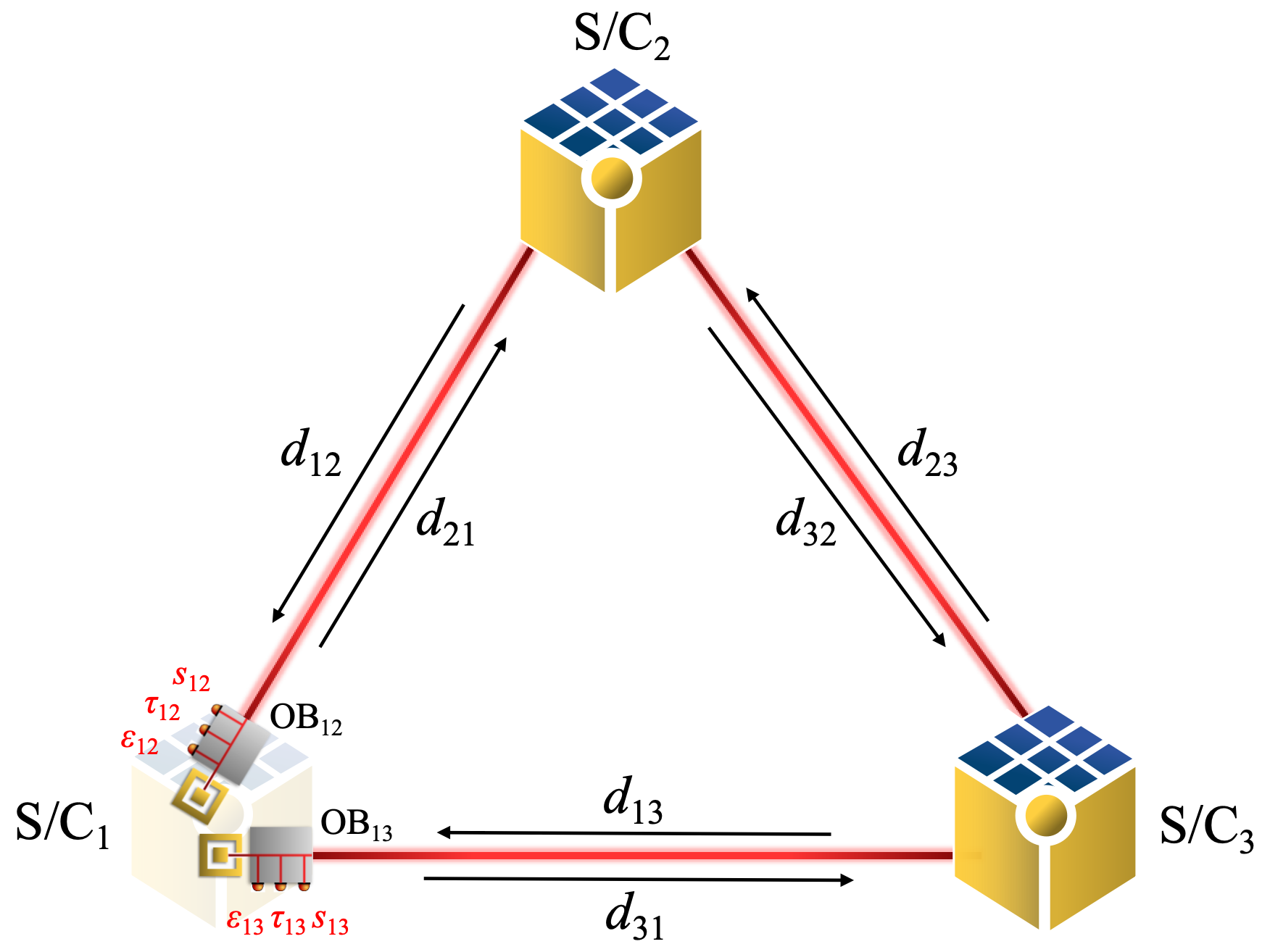}
    \caption{A schematic layout of  Taiji's constellation. Each S/C is equipped with two OBs, on which  the readouts of science, reference, and test-mass interferometer  are generated,  including the carriers  and sidebands. The indices of signals and noises are the same as the OBs hosting them. S/C$_2$ and S/C$_3$ share the same design of optical measurement system as S/C$_1$. For clarity and conciseness, the specific arrangements of optical components on the OBs are omitted.}
    \label{fig:taiji}
\end{figure}

The delay operators associated with $d^t_{ij}(\tau)$, $\hat{d}^{\hat{\tau}_i}_{ij}(\tau)$ and $\bar{d}^{\hat{\tau}_i}_{ij}(\tau)$ are denoted as $\textbf{D}_{ij}$, $\textbf{D}_{ij}^{\hat{\tau}_i}$and $\bar{\textbf{D}}_{ij}^{\hat{\tau}_i}$, respectively.
Assuming $f$ is a phase signal generated  on S/C$_j$, after propagated to S/C$_i$ and timestamped by the clock onboard S/C$_i$ as $\tau$, the resulting signal can be described equivalently with the two expressions below~\cite{TDINoSync}:
\begin{equation}\label{eq:DtoDhat}
    \textbf{D}_{ij}f^t(t)|_{t=t^{\hat{\tau}_i}(\tau)} = \textbf{D}_{ij}^{\hat{\tau}_i} f^{\hat{\tau}_j}(\tau).
\end{equation}
The left hand side is expressed in TCB, with the arriving TCB time $t=t^{\hat{\tau}_i}(\tau)$ determined according to Eq.~(\ref{eq:time_frame_conversion}), and the right hand side is expressed in the clock time of S/C$_i$. The equivalence holds since they describe the same event.
On the other hand, according to the definition of $\hat{d}_{ij}^{\hat{\tau}_i}(\tau)$, the difference between $\textbf{D}_{ij}^{\hat{\tau}_i} f^{\hat{\tau}_j}(\tau)$ and $\bar{\textbf{D}}_{ij}^{\hat{\tau}_i} f^{\hat{\tau}_j}(\tau)$ equals $f$'s time derivative multiplying $\hat{d}^{\hat{\tau}_i}_{ij}(\tau) - \bar{d}^{\hat{\tau}_i}_{ij}(\tau)$ (i.e. the clock jitter term), which, in the case where $f$ represents some instrumental noise, is a second-order noise term and can be neglected. Therefore,  we get a relationship between the delayed noises:
\begin{equation}\label{eq:DhattoDbarhat}
    \textbf{D}_{ij}^{\hat{\tau}_i} f^{\hat{\tau}_j}(\tau) \approx \bar{\textbf{D}}_{ij}^{\hat{\tau}_i} f^{\hat{\tau}_j}(\tau).
\end{equation}

In this paper, we construct   TDI combinations  from detrended data, namely the large phase ramps corresponding to the slow-varying MHz beatnote frequencies have already been  removed with methods such as polynomial fitting or bandpass filtering~\cite{Olaf_PhD_thesis,clock_jitter_reduction}, and the only terms left are the ones representing noises and GW signals.
The  beatnote signals  are usually modeled in terms of LTT (e.g. Ref.~\cite{Heinzel_2011,TintoClockNoise,Bayesian_TDIR1,Bayesian_TDIR2}). 
Working on the detrended data enables us to recast their expressions in terms of TDIR using  Eq.~(\ref{eq:DtoDhat}) and Eq.~(\ref{eq:DhattoDbarhat}):
\begin{eqnarray}
    s_{ij, c}^{\hat{\tau}_i} &=& \bar{\textbf{D}}^{\hat{\tau}_i}_{ij}p^{\hat{\tau}_j}_{ji} -  p^{\hat{\tau}_i}_{ij} 
    +  \bar{\textbf{D}}^{\hat{\tau}_i}_{ij}\Delta_{ji}^{\hat{\tau}_j} + \Delta_{ij}^{\hat{\tau}_i}
    - a^{\hat{\tau}_i}_{ij, c} q_i^{\hat{\tau}_i} \label{eq:sci_ifo}\nonumber \\ 
    && + N^{s, c, {\hat{\tau}_i}}_{ij},  \\
    s_{ij, sb}^{\hat{\tau}_i} &=& \bar{\textbf{D}}^{\hat{\tau}_i}_{ij}p^{\hat{\tau}_j}_{ji} -  p^{\hat{\tau}_i}_{ij} 
    +  \bar{\textbf{D}}^{\hat{\tau}_i}_{ij}\Delta_{ji}^{\hat{\tau}_j} + \Delta_{ij}^{\hat{\tau}_i}
    - a^{\hat{\tau}_i}_{ij, sb} q_i^{\hat{\tau}_i} \nonumber \\
    && + \nu^m_{ji}\bar{\textbf{D}}^{\hat{\tau}_i}_{ij}q_j^{\hat{\tau}_j} - \nu^m_{ij} q_i^{\hat{\tau}_i} 
    +  N^{s, sb, {\hat{\tau}_i}}_{ij},  \\
    \tau_{ij, c}^{\hat{\tau}_i} &=& p^{\hat{\tau}_i}_{ik} -  p^{\hat{\tau}_i}_{ij} - b^{\hat{\tau}_i}_{ij, c} q_i^{\hat{\tau}_i} + N^{\tau, c, {\hat{\tau}_i}}_{ij},  \\
    \varepsilon_{ij, c}^{\hat{\tau}_i} &=& p^{\hat{\tau}_i}_{ik} -  p^{\hat{\tau}_i}_{ij} 
    + 2 \left(\Delta_{ij}^{\hat{\tau}_i} - \delta^{{\hat{\tau}_i}}_{ij}\right)
    - b^{\hat{\tau}_i}_{ij, c} q_i^{\hat{\tau}_i}  \nonumber \\ 
    && + N^{\varepsilon, c, {\hat{\tau}_i}}_{ij} \label{eq:tm_ifo},
\end{eqnarray}
where $s_{ij}^{\hat{\tau}_i}, \tau_{ij}^{\hat{\tau}_i}, \varepsilon_{ij}^{\hat{\tau}_i}$ represent the science (inter-spacecraft), reference and test-mass interferometor readouts in the phase unit. Subscript ``c''/``sb'' is short for the carrier/sideband, while $ij \in \{12, 21, 23, 32, 31, 13\}$ stands for the index of optical bench where the measurements are taken. 
On the r.h.s of the equations, $p$, $\Delta$, $\delta$, $q$, $N^{\rm ifo}$ denote the laser phase fluctuation caused by LFN, the phase noises originating from  the motions of optical benches and test masses, the clock jitter,  and  the  OMS noise of corresponding interferometer, respectively.  Besides, 
$\nu^m_{ij}$ is the 2.4/2.401GHz modulation frequency of sideband, and $a$, $b$ are the MHz beatnote frequencies of corresponding signals, which couple with the clock jitters to form the clock noise.  
The indices of signals are the same as their corresponding OBs. See FIG.~\ref{fig:taiji} for a schematic layout of  Taiji's constellation.


As is explained in Sec.~\ref{sec:introduction}, for the applications of TDI, it is crucial that the delays used to shift the signals must be consistent with 
the delays appearing in the equations of signals. 
Formally Eq.~(\ref{eq:sci_ifo}-\ref{eq:tm_ifo}) looks similar to  their ``LTT versions''~\cite{Heinzel_2011,TintoClockNoise,Bayesian_TDIR1,Bayesian_TDIR2}, thus all the deduction and algorithm of TDI still applies, and the only two alterations which should be made are to change the delays from LTT to TDIR, and to interpret the beatnote signals as the unsynchronized  measurements. We therefore proved that TDIR can be used  to suppress LFN on unsynchronized data.

{\color{black} By making slight modifications to the standard formulae in the literature~\cite{Heinzel_2011,TintoClockNoise,Bayesian_TDIR1,Bayesian_TDIR2}}, we explicitly present the expressions for the complete TDI data processing flow using TDIR.
Firstly, the test-mass-to-test-mass interferometry (denoted as intermediate variable $\xi$) is constructed to eliminate the OB noises:
\begin{equation}\label{eq:intervar_xi}
    \xi_{ij}^{\hat{\tau}_i} = s_{ij}^{\hat{\tau}_i} + \frac{\tau_{ij}^{\hat{\tau}_i} - \varepsilon_{ij}^{\hat{\tau}_i}}{2} + \bar{\textbf{D}}_{ij}^{\hat{\tau}_i}\frac{\tau_{ji}^{\hat{\tau}_j} - \varepsilon_{ji}^{\hat{\tau}_j}}{2},
\end{equation}
with $ij \in \{12, 21, 23, 32, 31, 13\}$.
Secondly, intermediate variable $\eta$ is synthesized to eliminate half of the LFN on each S/C:
\begin{equation}\label{eq:intervar_eta}
    \eta_{ij}^{\hat{\tau}_i} = \xi_{ij}^{\hat{\tau}_i} + \bar{\textbf{D}}^{\hat{\tau}_i}_{ij}\frac{\tau_{ji}^{\hat{\tau}_j} - \tau_{jk}^{\hat{\tau}_j}}{2}, \quad
    \eta_{ik}^{\hat{\tau}_i} = \xi_{ik}^{\hat{\tau}_i} + \frac{\tau_{ij}^{\hat{\tau}_i} - \tau_{ik}^{\hat{\tau}_i}}{2},
\end{equation}
where $\{i, j, k\} \in \{\{1, 2, 3\}, \{2, 3, 1\}, \{3, 1, 2\}\}$.
The third step is the combination of TDI channels. We take the second-generation Michelson $X_2$ channel as an example:
\begin{eqnarray}\label{eq:X2_channel}
    X_2^{\hat{\tau}_1} &=& \left(1 - \bar{\textbf{D}}^{\hat{\tau}_1}_{121} - \bar{\textbf{D}}^{\hat{\tau}_1}_{12131} + \bar{\textbf{D}}^{\hat{\tau}_1}_{1312121}\right) \left(\eta_{13}^{\hat{\tau}_1}+\bar{\textbf{D}}^{\hat{\tau}_1}_{13}\eta_{31}^{\hat{\tau}_3}\right) \nonumber \\ 
    &-& \left(1 - \bar{\textbf{D}}^{\hat{\tau}_1}_{131} - \bar{\textbf{D}}^{\hat{\tau}_1}_{13121} + \bar{\textbf{D}}^{\hat{\tau}_1}_{1213131}\right) 
\left(\eta_{12}^{\hat{\tau}_1}+\bar{\textbf{D}}^{\hat{\tau}_1}_{12}\eta_{21}^{\hat{\tau}_2}\right), \nonumber \\
\end{eqnarray}
and the $Y_2$ and $Z_2$ channels can be obtained by cyclical permutation of the indices : $1 \rightarrow 2$, $2 \rightarrow 3$, $3\rightarrow 1$.
The time frame of each TDI channel will be the clock time of the ending S/C of the virtual optical path (e.g. S/C$_1$ for $X_2$), since the conversion of time frames from the emitting S/C to the receiving S/C is also encoded in the TDIR (i.e. the slow-varying part of  Eq.~(\ref{eq:mpr_t}-\ref{eq:mpr_tau}) contains the clock deviations of both S/Cs). 
At last, the residual clock noise (originating from the $q$ terms in Eq.~(\ref{eq:sci_ifo}-\ref{eq:tm_ifo})) in the second-generation TDI channels is still 2 - 3 orders larger than the desired noise level, thus an extra step is required to mitigate this noise source. Various  clock noise reduction algorithms exist in the literature~\cite{TintoClockNoise,clock_jitter_reduction,Tianqin_clock_noise}. Here we represent this step in a unified abstract form:
\begin{equation}
    X_{2, c}^{\hat{\tau}_1} = X_{2}^{\hat{\tau}_1} - X_{2, q}^{\hat{\tau}_1},
\end{equation}
$X_{2, q}$ being the clock noise correction term.
This step is a necessity for TDIR, or otherwise the TDI data streams would be dominated by the clock noise, 
making the noise modeling  in the TDIR likelihood (using the PSD of secondary noises, see Sec.~\ref{sec:Bayesian}) incorrect. Therefore, we suggest that the practical implementation of TDIR should encompass this crucial step. For brevity we will drop the subscript ``$c$'' in the following, and one should keep in mind that the clock noise reduction step should be added to the TDI data processing flow.

It is worth noting that we have not set any restriction  to the physical origin of clock deviation $\delta \hat{\tau}_i^t(\tau)$. In principle, it can be 
either the original (uncorrected) clock deviations, 
or the residual timestamp errors left by a preliminary synchronization using  ground-tracking data (ground-tracking synchronization hereafter), whose  uncertainty is reported to be  $\sim 0.1 \ {\rm ms}$~\cite{ranging_fusion}. 
In this paper we explore the application of TDIR in the second scenario.
Placing this preliminary synchronization step prior to TDI offers the opportunity to  combine different TDI channels, so as to achieve better constraint on the delay parameters.  

Besides, it is reported by Ref.~\cite{ranging_fusion}  that the orbital determination provided by  ground tracking enables us to  calculate LTTs~\cite{OD_LTT} (ground-tracking ranges hereafter) with an uncertainty of $50 \ {\rm km}$  ($\sim 0.1$ ms). This information can be used to model the secondary noises in the calculation of TDIR likelihood (see Sec.~\ref{sec:Bayesian}), and we will demonstrate that it allows to accelerate the processing of clock noise in TDIR.
The differences between ground-tracking ranges and fiducial TDIR are  $\delta \hat{\tau}_i^t(\tau) - \delta \hat{\tau}_j^t\left[\tau - d^t_{ij}(\tau)\right]$ plus the aforementioned uncertainties of ground-tracking ranges, 
adding up to  total differences of less than 1 ms. 

To give a more intuitive explanation of the principles, beatnote signals in the above derivation are all represented  in phase units. Following recent progress in the TDI algorithm~\cite{FrequencyTDI}, we perform  TDI processing with simulated data in the frequency unit. Accordingly, all the time delay operators should be replaced by the Doppler delay operators:
\begin{equation}
    \dot{\bar{\textbf{D}}}_{ij}^{\hat{\tau}_i} f = \left(1- \dot{\bar{d}}_{ij}^{\hat{\tau}_i}(\tau)\right)\bar{\textbf{D}}_{ij}^{\hat{\tau}_i} f.
\end{equation}

\subsection{Requirements of noise suppression on TDIR}\label{subsec:requirement}

Throughout  the entire  TDI data processing flow, sufficient  noise suppression demands the delay operators used in TDI combination to be coincident with the ones appearing in the expressions of signals. 
The  error of TDIR will couple with instrumental noises and result in undesirable noise residuals in the TDI combinations. 
In this subsection, The coupling of ranging error with LFN, OB noise and clock noise will be investigated through theoretical analysis and simulation. These discussions  can be used to set a threshold  for the accuracy of TDIR.

The noise cancellation performance of TDI is usually evaluated  relative to the   secondary noises,  mainly including the  acceleration noise of test masses, corresponding to the $\delta$ terms in  Eq.~(\ref{eq:sci_ifo}-\ref{eq:tm_ifo}), and the OMS noises, corresponding to the $N^{\rm ifo}$ terms. To make an order-of-magnitude analysis, we take the $X_2$ channel as an example, and consider  the equal-arm case. The PSD of secondary noises  reads
{\color{black}\begin{eqnarray}\label{eq:TDIX2ndnoise}
    P^{\rm 2nd}_{X_2}(f) &=& 64\sin^2(2u)\sin^2(u) \nonumber \\ 
    &&\times \left[P_{\rm OMS}(f) + P_{\rm ACC}(f)(\cos 2 u + 3)\right],
\end{eqnarray}}
where $u\equiv 2\pi f L / c$,  $L=3\times 10^9 \ {\rm m}$ being the nominal arm-length of Taiji.  The instrumental noise models used in the analyses  and simulations are detailed in Appendix~\ref{sec:appendix}. This expression also applies to the $Y_2$ an $Z_2$ channels. We set $1/10 P^{\rm 2nd}_{X_2}(f)$ as the thresholds for the noise cancellation capability of TDI.

\begin{figure}
    \centering
    \includegraphics[width=0.5\textwidth]{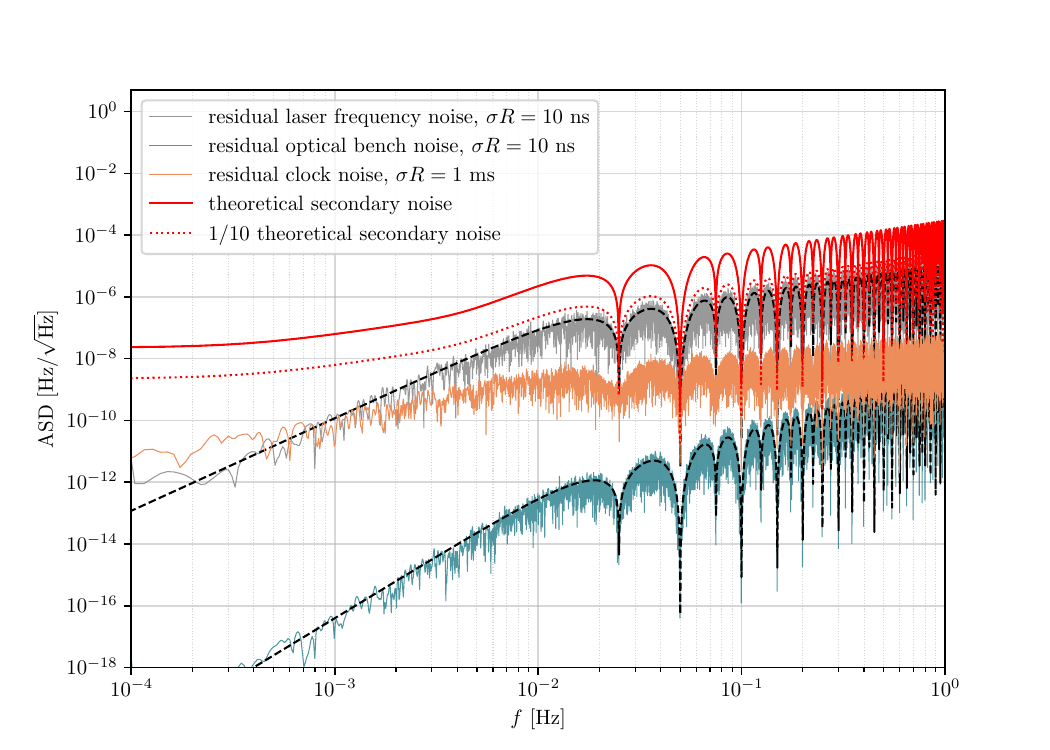}
    \caption{Analytical and simulation results of the residual noises caused by  ranging errors in the Michelson $X_2$ channel. All the ranging errors are generated from normal distributions $\mathcal{N}(0, \sigma R)$, with the values of $\sigma R$ vary for different noises. The grey, green and orange ASD plots represent the simulated residual LFN, OB noise and clock noise,  with the corresponding theoretical results shown in  black dashed curves. The red solid and  dotted lines stand for  1 and 1/10 times  the secondary noises, respectively, which are set as the thresholds for TDI noise suppression.}
    \label{fig:requirements_of_OB_clock_laser}
\end{figure}

Next, we  investigate the residual {\color{black}LFN}, OB noise and clock noise  caused by the ranging errors. 
For  generality, we do not set a specific laser locking scheme~\cite{locking,Olaf_PhD_thesis} but rather assume all the six lasers are independent. 
Besides, to derive analytical expressions for a given level of  ranging accuracy, the  errors of ranging are set as  constant biases dubbed $\Delta R_{ij}$.
According to Eq.~(\ref{eq:intervar_xi}-\ref{eq:X2_channel}), the PSDs of the  residual LFN and OB noise  in the $X_2$ channel are 
\begin{eqnarray}
    P^{\rm L}_{X_2}(f; \Delta R_{ij}) &=& 64\sin^2(2u)\sin^2(u) \omega^2 \nonumber \\
    &&\times P_{\rm L}(f) \Delta R^2, \label{eq:laser_residual}\\
    P^{\rm OB}_{X_2}(f; \Delta R_{ij}) &=&  64 \nu_0^2 \sin^2(2u)\sin^2(u) \omega^4 \nonumber \\ 
    &&\times P_{\rm OB}(f) \Delta R^2, \label{eq:OB_residual}
\end{eqnarray}
where $\omega \equiv 2\pi f$,   $\nu_0 = 281.6 \ {\rm THz}$ is the central frequency of laser, and we have defined
\begin{equation}\label{eq:avg_bias}
    \Delta R \equiv \sqrt{\frac{\Delta R_{12}^2+\Delta R_{21}^2+\Delta R_{13}^2+\Delta R_{31}^2}{4}}
\end{equation}
as the average  ranging error.  
Eq.~(\ref{eq:laser_residual}) is in consistency with the results of previous studies~\cite{TDINoSync,laser_residual}, and one can easily  verify that a $\mathcal{O}(10) \ {\rm ns}$ requirement is reasonable for the noise models in consideration. While, under the same $\Delta R$, the residual OB noise is much smaller. 
The algorithms of clock noise reduction varies in the literature (e.g.~\cite{TintoClockNoise,clock_jitter_reduction,Tianqin_clock_noise}), thus the residual clock noise  can not be written in a unified analytical form. We adopt the treatment of Ref.~\cite{clock_jitter_reduction}, and explore the impact of ranging error  via a clock-noise-only simulation.

\begin{figure}
    \centering
    \includegraphics[width=0.5\textwidth]{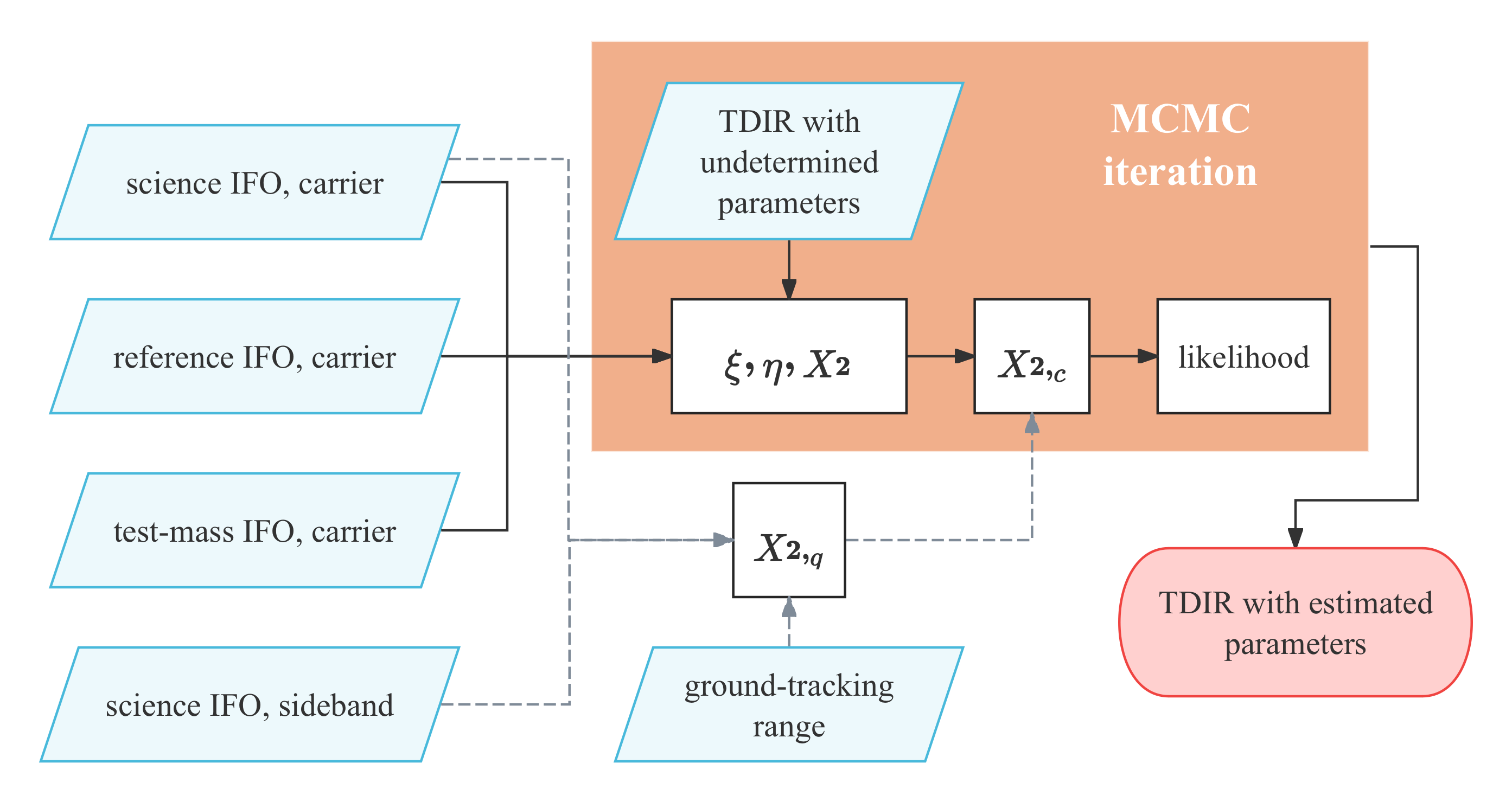}
    \caption{{\color{black}A simplified flowchart for TDIR. Shown with the light blue parallelogram boxes are the main inputs of the algorithm. The steps in the orange box depend on the undetermined TDIR parameters and should be re-calculated in each iteration of of the MCMC sampling. 
    The dashed arrows represent the clock noise processing steps. As the suppression of clock noise  has lower requirements on the accuracy of ranging, the correction term $X_{2, q}$ can be pre-calculated with the ground-tracking ranges,  and hence doesn't need to be included into the loop. This effectively reduces the total computation time of TDIR.}}
    \label{fig:FlowChart}
\end{figure}

The analytical and simulation results  are visualized in FIG.~\ref{fig:requirements_of_OB_clock_laser}. In our simulation, the ranging errors are generated according to  normal distributions $\mathcal{N}(0, \sigma R)$, with the values of $\sigma R$ vary for different noises, which are 10 ns, 10 ns, 1 ms for the LFN, OB noise and clock noise, respectively.
The grey, green and orange amplitude spectral density (ASD) plots represent the simulated residual LFN, OB noise and clock noise,  with the corresponding theoretical results shown in  black dashed curves. 
The red solid and  dotted lines stand for  1 and 1/10 times  the secondary noises, respectively, which are set as the thresholds for TDI noise suppression.  
Besides the consistency between theoretical analysis and simulations, the figure also indicates  that although the  ground-tracking ranges deviate from the fiducial TDIR for less than 1 ms, they are  already sufficient for  suppressing the clock noises. 
This result is  beneficial for the practical implementation of TDIR. Currently the post-TDI clock noise algorithms are usually more complex than TDI itself, therefore require more computational time. To tackle this issue, we do not need to compute $X_{2, q}$ at every iteration of the MCMC run, but rather calculate it with the ground-tracking ranges in advance. {\color{black} Based on this idea, we have designed the whole algorithm of  TDIR, which is shown with a simplified flowchart in FIG.~\ref{fig:FlowChart}.}
Furthermore, within the regions of parameter space  of our interest (tens of nanoseconds around the fiducial TDIR), OB and clock noises only make negligible  contributions to the total noise budget, thus the performance of TDIR is mainly determined by LFN and the secondary noises.

\begin{figure*}
\includegraphics[width=0.48\textwidth]{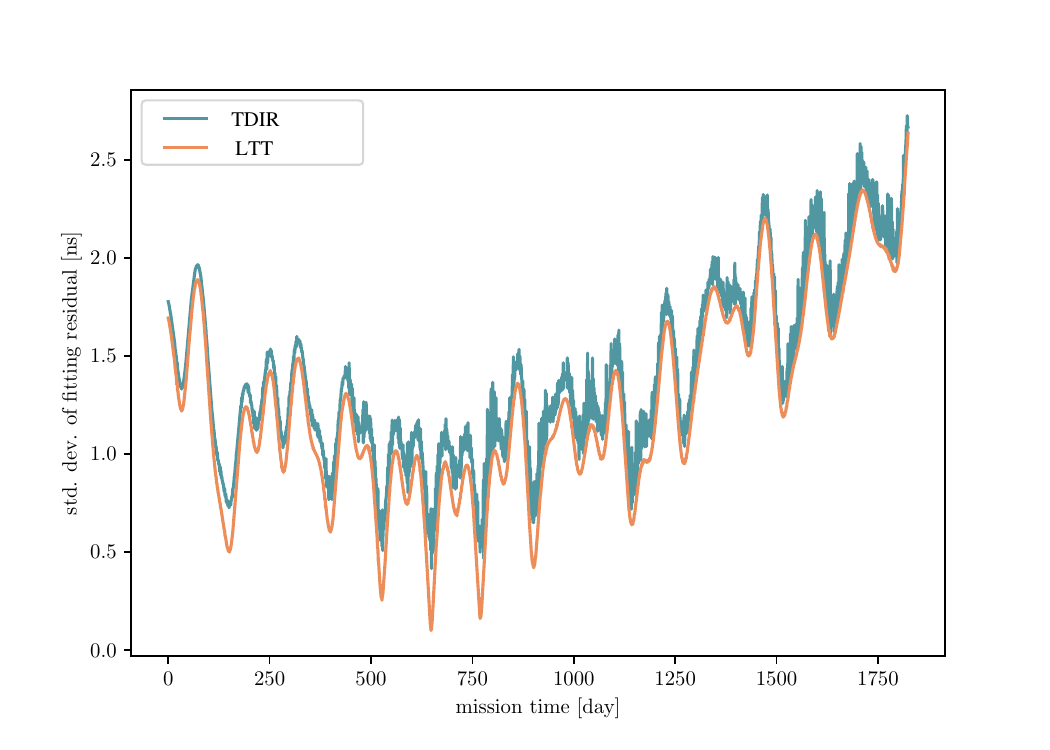} 
\includegraphics[width=0.48\textwidth]{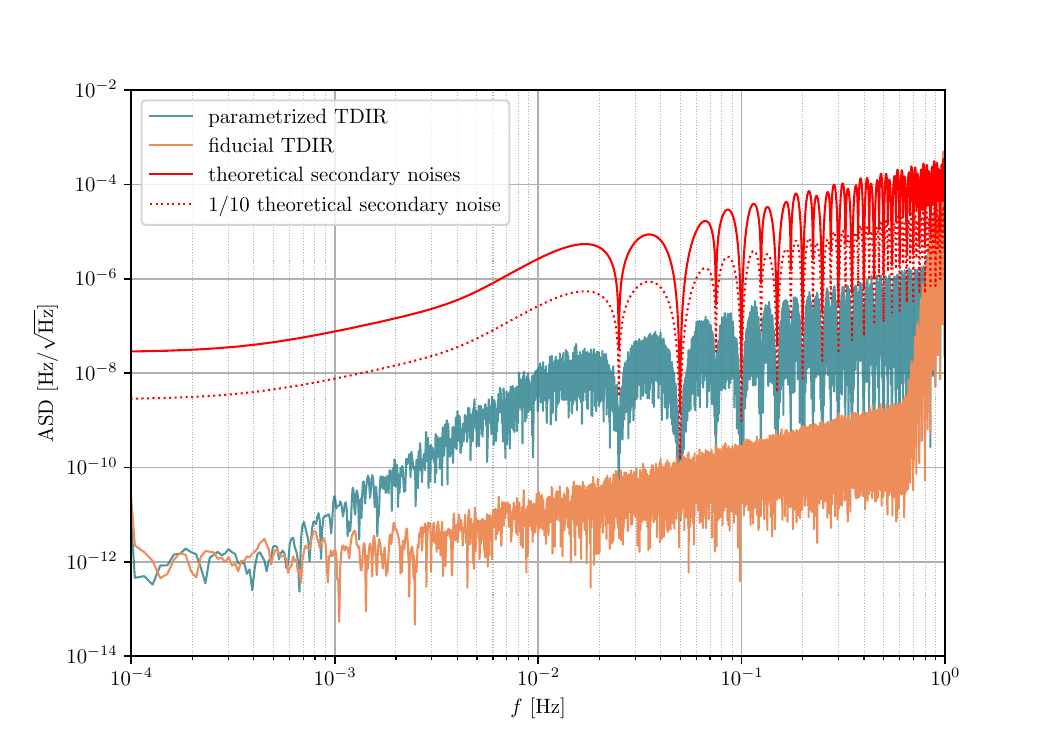}
\caption{\label{fig:bias_stat} Left panel: the standard deviations of the differences between the fitting  and  true LTTs (orange curve) /  TDIRs (green curve), evaluated per day in the whole mission period.  The TDIRs are calculated using on the original (uncorrected) clock deviations. Right panel: the ASDs of $X_2$ channels constructed from a LFN-only simulation using  the fiducial and parametrized TDIRs. The simulation is conducted based on the 1st day of Taiji's numerical orbit, and the uncorrected clock model. All the parameters are determined via  least square fit.}
\end{figure*}

\subsection{Parametrization of TDIR}

The TDIR algorithm  utilizes parameterized function as the hypothetical model, and the parameters will  be constrained using  the beatnote signals. 
Ref.~\cite{Bayesian_TDIR2} proposed a fitting form for the delays based on Keplerian orbit. This approach,  according to our analysis, is only applicable when the delays are LTTs. 
Applying TDI to unsynchronized data demands that the delays also include information about clock deviations, hence a more general parametrization is desired. 
We employ a quadratic polynomial parametrization to fit the delays over   a day:
\begin{equation}\label{eq:parametrization}
    R_{ij}(\tau; \bm{\theta}) = A_{ij}\tau^2 + B_{ij}\tau + C_{ij},
\end{equation}
where $\bm{\theta} = \{A_{ij}, B_{ij}, C_{ij}\}, ij \in \{12, 21, 23, 32, 31, 13\}$ includes 18 parameters in total.

The degree of polynomial  is closely related to the time duration of data.  Due to the complexity of orbital dynamics and relativistic corrections, longer duration naturally requires higher-degree polynomials and more model parameters, hence heightening the challenge of MCMC parameter estimation. On the other hand, as  will be analyzed in Sec.~\ref{sec:Bayesian}, achieving better constraint on the  parameters  requires the availability of more data points.
After balancing between difficulty and precision, we choose the duration of data to be one day.

Through calculations  based on the numerical orbit, we find that quadratic polynomial is sufficient for this timescale.
The numerical orbit  is provided by Ref.~\cite{Orbit,Orbit_github}, which takes into consideration the influence of the
gravitational fields of celestial bodies and is qualified for space-based GW detection.
Throughout the whole mission period of 5 years,  we compute the LTTs using the formulas in Ref.~\cite{relativistic_link}, and fit Eq.~(\ref{eq:parametrization}) within each day using the least square fitting method. The standard deviations of the differences between the fitting and  true  LTTs are shown with the orange curve in  the left panel of FIG.~\ref{fig:bias_stat}. 
Then, the original (uncorrected) clock deviations are set according to the USO model given in Ref.~\cite{TDINoSync}, which assumes a $5\times 10^{-7} \ {\rm s/s}$ drift (43 ms/day). We repeat above calculations to the fiducial TDIR resulting from this clock model, and the outcome is represented by the green curve in the same panel. 
For TDIR corresponding to the residual clock deviation left by ground-tracking synchronization, the result lies somewhere between the green and orange curves.  According to the threshold established in Sec.~\ref{subsec:requirement}, we justify Eq.~(\ref{eq:parametrization}) to be a reasonable parametrization. 
The right panel of FIG.~\ref{fig:bias_stat} shows the ASDs of $X_2$ channels constructed from a LFN-only simulation using  the fiducial and parametrized TDIRs, with the parameters determined via  least square fit.
The simulation is conducted based on the 1st day of Taiji's numerical orbit, and the uncorrected clock model. 
It is obviously shown that 
the result of fiducial TDIR manifests  a very ideal effect of  LFN suppression. 
The residual LFN resulting from  parametrized TDIR is well below 1/10 of the secondary noises, 
although larger than the  fiducial one. 

\section{TDIR in the Bayesian framework}\label{sec:Bayesian}
\begin{figure*}
    \centering
    \includegraphics[width=0.48\textwidth]{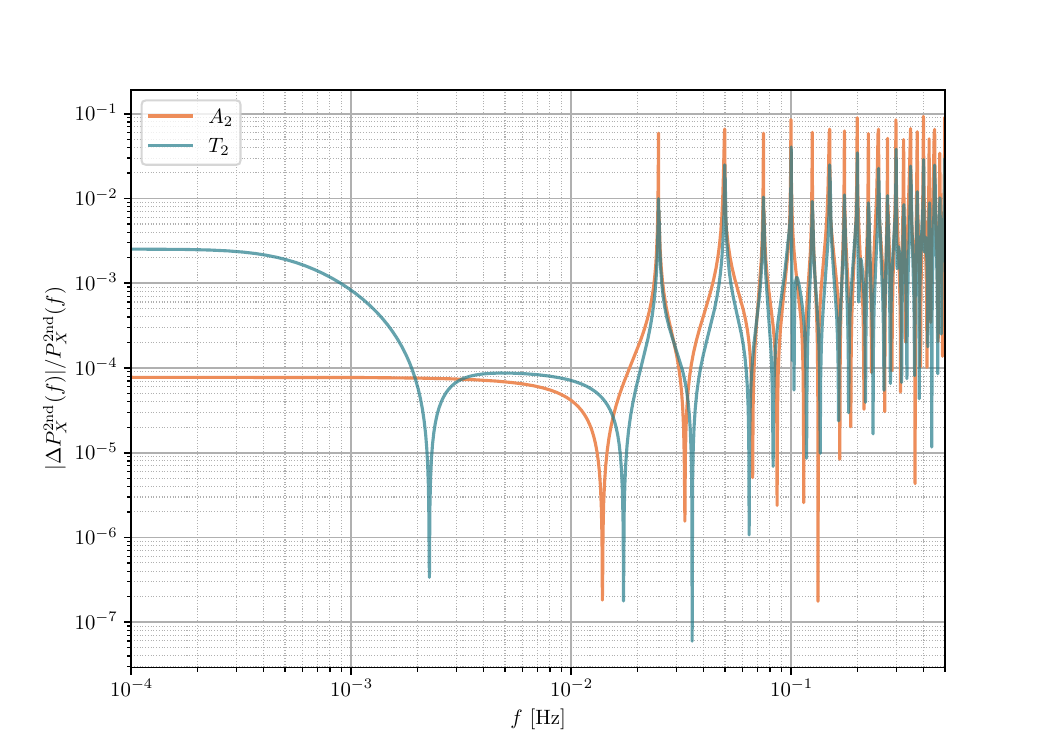} 
    \includegraphics[width=0.48\textwidth]{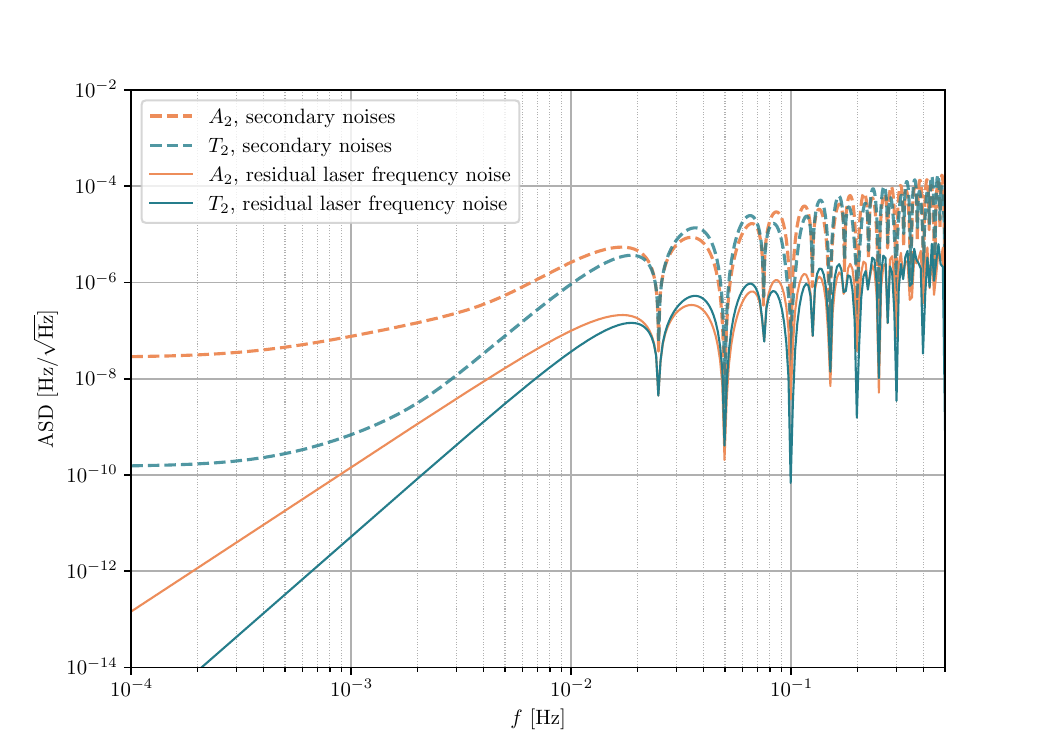} \\
    \caption{{\color{black}Left panel: The relative differences of the theoretical secondary noise PSDs caused by the variations of armlengths at 1 ms order. Right panel: The ASDs of  secondary noises and residual LFN in the $A_2$ and $T_2$ channels, with the armlengths calculated from the orbit data in the 1st day, and the ranging error are set to 10 ns. For both panels the amplitudes of $E_2$ are similar to those of $A_2$ hence not shown. }}
    \label{fig:AT_2nd_vs_laser}
\end{figure*}

The Bayesian statistical inference framework is a mathematical  approach to reasoning about uncertainty, which allows for the incorporation of prior knowledge and noisy data into posterior distributions. According to the Bayes' theorem, the posterior of model parameter reads
\begin{equation}\label{eq:posterior}
    p\left(\bm{\theta}|d\right) = \frac{\mathcal{L}(d|\bm{\theta})\pi(\bm{\theta})}{p(d)},
\end{equation}
where $\mathcal{L}(d|\bm{\theta})$ is the likelihood function, namely the  distribution of observational data $d$ conditioned on the model parameter $\bm{\theta}$. Without any prior knowledge on the parameters,  the prior $\pi(\bm{\theta})$ is set as a multidimensional uniform distribution.
The evidence $p(d)$ only acts as a normalization factor  and is irrelevant to the values of parameters once the hypothetical model has been  chosen, thus 
\begin{equation}
    p(\bm{\theta}|d) \propto \mathcal{L}(d|\bm{\theta}).
\end{equation}
Assuming that the instrumental noises  are Gaussian and stationary, $\mathcal{L}$ can be expressed as:
\begin{eqnarray}\label{eq:likelihood}
    {\rm ln} \mathcal{L} &=& -\frac{1}{2}\left(X(\bm{\theta})|X(\bm{\theta})\right) = -2\int_{f_{\rm min}}^{f_{\rm max}} \frac{|\tilde{X}(f;\bm{\theta})|^2}{P^{\rm 2nd}_X(f)}df \nonumber \\
    &=& -2\sum_{f_k} \frac{|\tilde{X}(f_k;\bm{\theta})|^2\Delta f}{P^{\rm 2nd}_X(f_k)} = -\sum_{f_k} \frac{P_X(f_k;\bm{\theta})}{P^{\rm 2nd}_X(f_k)},
\end{eqnarray}
where $\tilde{X}(f; \bm{\theta})$ represents the Fourier transform of an arbitrary TDI channel named $X$, and $P_X(f;\bm{\theta})$ is its one-sided PSD (periodogram). The  data is whitten by  the theoretical one-sided PSD of the secondary noises  $P^{\rm 2nd}_X(f)$, since secondary noises dominate the TDI data  in the region of parameter space of our interest (tens of nanoseconds around fiducial TDIR). 
{\color{black} In the realistic unequal-arm situation, $P^{\rm 2nd}_X(f)$ also depends on the LTTs, which can take the average values of ground-tracking ranges within a day.
Using Eq.~(\ref{eq:X2_unequal}-\ref{eq:T2_secondary}), it can be verified  that neither the ranging uncertainties ($\sim 0.1 \ {\rm ms}$) nor the variations of LTTs during a day ($\sim 1 \ {\rm ms}$) have considerable impacts on  $P^{\rm 2nd}_X(f)$  via numerical or analytical calculation. 
Shown in the left panel of FIG.~\ref{fig:AT_2nd_vs_laser} are the relative differences of $P^{\rm 2nd}_{X}(f) \ (X \in \{A_2, E_2\})$ caused by the variations of armlengths at 1 ms order. As can be seen from the figure, the accuracy of modeling the secondary noise using the average ground-tracking ranges is typically greater than 99\%, and even at the characteristic frequencies ($f = 0.025 n $ Hz) where the armlengths have the greatest impacts, the accuracy is still above 90\%.}
Eq.~(\ref{eq:likelihood}) is equivalent to the ones adopted by previous studies~\cite{Bayesian_TDIR1,Bayesian_TDIR2}.
In the more general ``global fit''~\cite{global_fit} scenario where GW signals exist, ${\rm ln}\mathcal{L}$ should be replaced by a more familiar form $-1/2(X-h|X-h)$, $h$ being the templates of GW signals.  

Simultaneously estimating the delay parameters and  GW source parameters  can 
be a challenging task. 
Additionally, Taiji will be able to detect the GWs from tens of millions of compact binaries in the Milky Way, and the majority of them cannot be resolved individually, resulting in a foreground confusion noise in the mHz band~\cite{GB}. Precise modeling of this noise is only possible when all the bright GW signals have been correctly recognized and subtracted.
Therefore, to reduce the encounters with GW signals and foreground noise,   
we set $f_{\rm min} = 10 \ {\rm mHz}$, above the most sensitive band of Taiji. This choice of $f_{\rm min}$ also allows us to accelerate the estimation process by incorporating a preliminary search phase, see Sec.~\ref{sec:Simulation}.
Ideally, under the   condition that the computing capacity is sufficient  or  the algorithm has been  optimized in speed, one may discard this  phase and raise $f_{\rm min}$ above 0.1 Hz to further minimize the impact of GWs.


In this paper, we perform the  time shift of signals with Lagrangian interpolation~\cite{PhysRevD.70.081101}. Interpolation error increases dramatically near the Nyquist frequency.  For a sampling frequency of 4 Hz, we find that Lagrangian interpolation of order 15 can enhance the efficiency of TDI calculation (compared to the widely adopted  31 order) with acceptable  compromise in precision. Under this configuration, interpolation error is well below the secondary noises when $f < f_{\rm max} = 0.5 \ {\rm Hz}$.

The Fisher information matrix  is widely employed to predict the uncertainty of parameter estimation in the context of  GW observations
~\cite{Fisher1,Fisher2}. 
We take the  equal-arm case as an example to roughly estimate  the precision   of the delays using the Fisher matrix formalism. 
For the simplified analysis, we only consider the Michelson $X_2$ channel and assume that the delay for each link  is represented by a constant $R_{ij}$. Under these simplifications, the logarithmic likelihood function can be rewritten as 
\begin{eqnarray}\label{eq:loglike_X2}
    {\rm ln}\mathcal{L} &=& -\sum_{f_k} \frac{P^{\rm L}_{X_2}(f_k;\bm{\theta}) +  P^{\rm 2nd}_{X_2}(f_k)}{P^{\rm 2nd}_{X_2}(f_k)} \nonumber \\
    &=& -\sum_{f_k} \left[1 + \frac{P^{\rm L}_{X_2}(f_k;\bm{\theta})  }{P^{\rm 2nd}_{X_2}(f_k)}\right].
\end{eqnarray}
When using this single channel, 
it can be  easily verified that there is no correlation between  different  arms via definition 
\begin{equation}\label{eq:correlation}
    {\rm Cov}_{\theta_a\theta_b} \equiv \frac{\Gamma^{-1}_{\theta_a\theta_b}}{\sqrt{\Gamma^{-1}_{\theta_a\theta_a}\Gamma^{-1}_{\theta_b\theta_b}}},
\end{equation}
where 
\begin{equation}\label{eq:Fisher}
    \Gamma_{\theta_a \theta_b} = - \frac{\partial^2 {\rm ln}\mathcal{L}}{\partial \theta_a \partial\theta_b}.
\end{equation}
is the element of Fisher matrix corresponding to parameters $\theta_a$ and $\theta_b$, and $\Gamma^{-1}$ denotes  the inverse matrix of $\Gamma$.
Therefore, the Cramer-Rao bounds for the uncertainties of delays are   
\begin{eqnarray}\label{eq:variance}
\sigma_{R_{ij}}^{-2} &=& \Gamma_{R_{ij} R_{ij}} = -\frac{\partial^2 {\rm ln} \mathcal{L}}{\partial R_{ij}^2} 
= \sum_{f_k} \frac{\frac{\partial^2 P^{\rm L}_{X_2}}{\partial R_{ij}^2}(f_k, \bm{\theta})}{P^{\rm 2nd}_{X_2}(f_k)} \nonumber \\
&\approx& \frac{c^2}{2\nu_0^2} \frac{A_{\rm L}^2}{A_{\rm OMS}^2} N, \quad ij \in \{12, 21, 13, 31\}.
\end{eqnarray}
The last equivalence holds due to the fact that OMS noise dominates the   noise budget at the frequency range of interest. Therefore 
\begin{equation}
    \sigma_{R_{ij}} \approx \frac{\sqrt{2}\nu_0}{c} \frac{A_{\rm OMS}}{A_{\rm L}} \frac{1}{\sqrt{N}} \approx \frac{3.54\times 10^{-7} \ {\rm s}}{\sqrt{N}}.
\end{equation}
For $f_{\rm min} = 10 \ {\rm mHz}$, $f_{\rm max} = 0.5 \ {\rm Hz}$ and $df = 1 / {\rm day}$, the number of data points in the Fourier domain is $N=42336$, thus $\sigma_{R_{ij}}\approx 1.72 \ {\rm ns}$. 
This represents a rather optimistic estimate,  as more complex parametrizations would likely introduce correlations between the parameters characterizing each arm, which could inflate the errors.

Data of a single TDI channel 
$X_2$ can only place constraints on $R_{12}$, $R_{21}$, $R_{13}$ and $R_{31}$. Using multiple channels is necessary for sampling the entire parameter space,  and it can also help to obtain tighter constraints on  the parameters. 
In further analysis based on the simulated data, to fully utilize the information from different  TDI channels, and to maintain consistency with the conventions of GW source parameter estimation so as to be  integrated into the global fit pipeline, we employ the ``optimal'' TDI channels constructed from the Michelson channels as~\cite{optimal_channels}:
\begin{eqnarray}
    A_2 &=& \frac{Z_2 - X_2}{\sqrt{2}},   \\ E_2 &=& \frac{X_2 - 2Y_2 + Z_2}{\sqrt{6}}, \\ T_2 &=& \frac{X_2 + Y_2 + Z_2}{\sqrt{3}},
\end{eqnarray}
and the total logarithmic likelihood function is the sum of $\{A_2, E_2, T_2\}$ channels. 
The PSDs of secondary noises in the optimal channels can be found in Appendix~\ref{sec:appendix}, where we have also presented the PSDs of residual LFN due to the coupling with ranging errors.
Observing Eq.~(\ref{eq:loglike_X2}), the contribution of a certain TDI channel to the total likelihood is determined by the ratio between residual LFN and secondary noises. These two terms for the $\{A_2, T_2\}$ channels are shown in {\color{black}the right panel of}  FIG.~\ref{fig:AT_2nd_vs_laser}, with the armlengths calculated from the numerical orbit data in the 1st day, and the ranging errors are set to 10 ns. The amplitudes of noises in $E_2$  are similar to those of $A_2$ hence not shown. As can be seen from the figure, within the frequency band in concern, each channel contributes similarly to the likelihood.

We know from the explanation in Sec.~\ref{sec:demonstration} that the timestamps of each Michelson channel is determined by the clock at the end of the virtual optical path. 
Some comments should be made about the combination of these ``roughly synchronized'' channels. 
Firstly, taking the $A_2$ channel as an example and assuming the desynchronization between S/C$_1$ and S/C$_3$ is a constant $\delta t$, then the error in  modeling $P_{A_2}^{\rm 2nd}$ with Eq.~(\ref{eq:A2_secondary})  is approximately $2 \omega \delta t {\rm Im}(Z_2X_2^*)$, much smaller than the other terms of $P_{A_2}^{\rm 2nd}$ under the condition of $\omega \delta t \ll 1$, which is met across the whole 0.1 mHz - 1 Hz band. Secondly, 
the  ``best fit'' delays resulting from  different TDI channels could mismatch due to the desynchronization of corresponding S/Cs. Ref.~\cite{UnequalArm} summarized the criteria for S/C movements to be  qualified  for space-based GW detection:  the relative velocities
between S/Cs should be smaller than 5 m/s for LISA and 6 m/s for Taiji. we take   a conservative value  of  10 m/s, which yields the variations of delays smaller than a nanosecond for $\delta t < 0.03 \ {\rm s}$.
Therefore, we conclude that a  0.1 ms order desynchronization  between different Michelson channels does not impact the combination into optimal channels.


\section{Simulation results}\label{sec:Simulation}
\subsection{Settings for data simulation and MCMC sampling}\label{subsec:settings}
Until now, the functional forms of residual clock errors left by ground-tracking synchronization for the Taiji mission are still under investigation. 
We treat the magnitudes reported in Ref.~\cite{ranging_fusion} as the state-of-the-art, and assume their form  as 
\begin{equation}\label{eq:residual_clock_model}
    \delta \hat{\tau}_i^t(\tau) = \sum_{m=0}^2 \mathcal{C}_{m, i} \tau^m, \quad i \in \{1, 2, 3\},
\end{equation}
with the coefficients generated from  normal distributions $\mathcal{C}_{m, i} \sim \mathcal{N}(0, 0.1 \ {\rm ms} / T_{\rm obs}^m)$, $T_{\rm obs} = 1 \ {\rm day}$. 
Each term contributes  0.1 ms to the total deviation. 
According to the theoretical analysis in Sec.~\ref{subsec:principle}, 
we do not expect  the specific shape of $\delta \hat{\tau}_i^t(\tau)$  to have a significant impact on the feasibility of TDIR. 
Regarding  parametrization Eq.~(\ref{eq:parametrization}), the left panel of FIG.~\ref{fig:bias_stat} shows that it is even applicable to the raw clock deviations, and the combination of different channels is valid up to the limits set at the end of Sec.~\ref{sec:Bayesian}.
Besides, we assume the ground-tracking range in each of the six laser links to deviate from corresponding LTT following the same form as Eq.~(\ref{eq:residual_clock_model}). 
Note that these coefficients  are independent of $\mathcal{C}_{m, i}$. 
These ranges are used as LTTs to  calculate the theoretical  PSDs  in the likelihood.
The  simulation of unsynchronized beatnote signals in the frequency unit follows the approach described in Ref.~\cite{LISA_instrument,Olaf_PhD_thesis}. 
To be specific, the signals are initially generated in the proper time of each S/C, then transferred  to TCB, and finally resampled  according to  the clock deviation  model set by Eq.~(\ref{eq:residual_clock_model}). 
To set a benchmark, we calculate the fiducial TDIR according to Eq.~(\ref{eq:mpr_t}-\ref{eq:mpr_tau}) and obtain the ``true values'' of  parameters via least square fitting.


We generate two sets of simulated data:
\begin{enumerate}
    \item [(1)] Simulation I: only contains the realization of LFN (labeled as ``only laser frequency noise'' in the figures);
    \item [(2)] Simulation II: contains the realizations of LFN as well as secondary noises (labelled as ``with secondary noises'' in the figures).
\end{enumerate}
The first simulation  is intended  to validate our parametrization and TDIR algorithm in an ideal situation, while the second one aims to test the performance of the algorithm under a more realistic scenario.
The simulations are all based on the 1st day of the numerical orbit, and the sampling frequency is set as 4 Hz.
All the instrumental noises are assumed to be Gaussian and stationary, and see Appendix~\ref{sec:appendix} for the noise models in the form of PSD. Still, for generality no specific laser locking scheme is set and all six lasers are independent.
Clock noise and OB noise are not present in either of the simulations since they do not contribute considerably to the likelihood, and the clock noise correction term can be calculated prior to TDIR, as is explained in Sec.~\ref{subsec:requirement}.

We use the affine-invariant MCMC ensemble sampler \texttt{emcee} to draw samples from the likelihood~\cite{emcee}. The stretch parameter is set to the default value $a=2$.   72 walkers (4 times the number of parameters) are employed to fully explore the shape of posterior.
The whole estimation process can be divided into two phases. The first phase is a preliminary  search, where we downsample the data to 1 Hz, and the interpolation order is set to 7  for the sake of efficiency. $f_{\rm max}$ is correspondingly reduced to 20 mHz to avoid interpolation error. Following the evaluation in Sec.~\ref{subsec:principle}, the preliminary  search starts from  1 ms intervals centered at the  ground-tracking ranges. Once the  preliminary  search reaches a steady posterior distribution,   the second phase, namely the refined search begins from the resultant $3 \ \sigma$ confidence intervals. This phase is based on the   original 4 Hz data,  and we adopt  the settings given in Sec.~\ref{sec:Bayesian}, i.e. interpolation order 15 and $f_{\rm max} = 0.5 \ {\rm Hz}$. 
For the refined search, we run at least 50 auto-correlation times after  burn-in  following the instruction of the code.

\subsection{Results of MCMC sampling}
\begin{figure*}
\includegraphics[width=\textwidth]{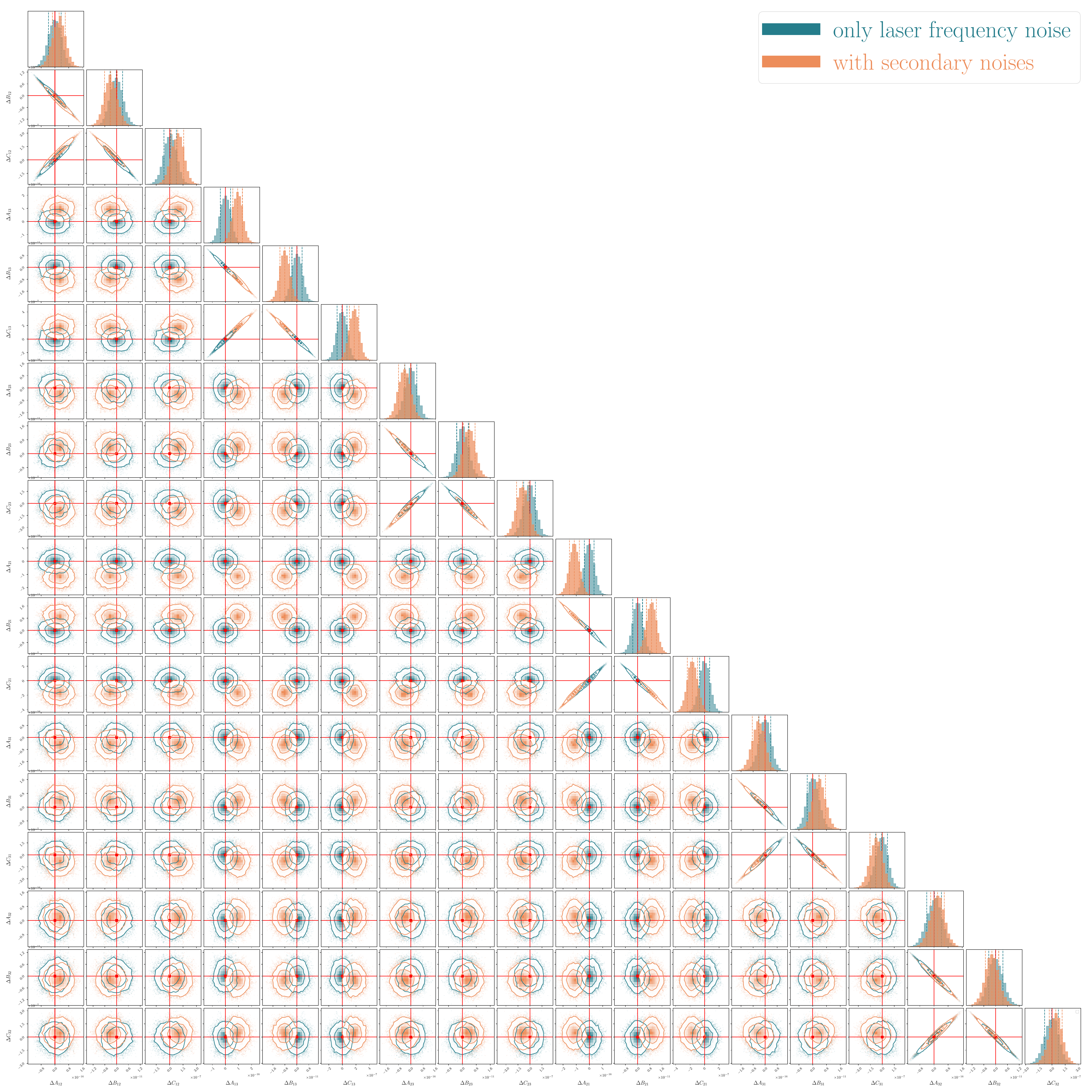}
\caption{\label{fig:posterior}Posterior distributions of the TDIR parameters. Results based on the LFN-only  and the  LFN + secondary noises are shown in green and orange corner plots, respectively, where the inner and outer contours represent the 1 $\sigma$ and 2 $\sigma$ ranges of the parameters, and the ``true values'' are marked with the red lines. In the 1D distribution plots on the diagonal, the vertical dashed lines indicate the 1 $\sigma$ ranges and median values of the corresponding parameter. }
\end{figure*}

 \begin{figure*}
    \centering
    \includegraphics[width=0.47\textwidth]{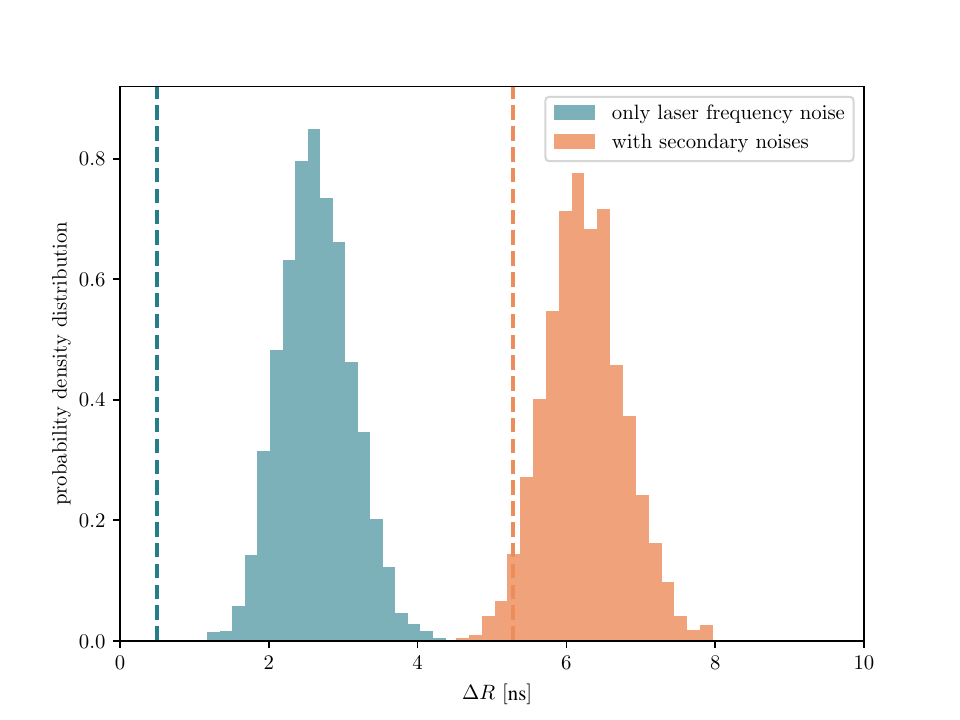}
    \includegraphics[width=0.49\textwidth]{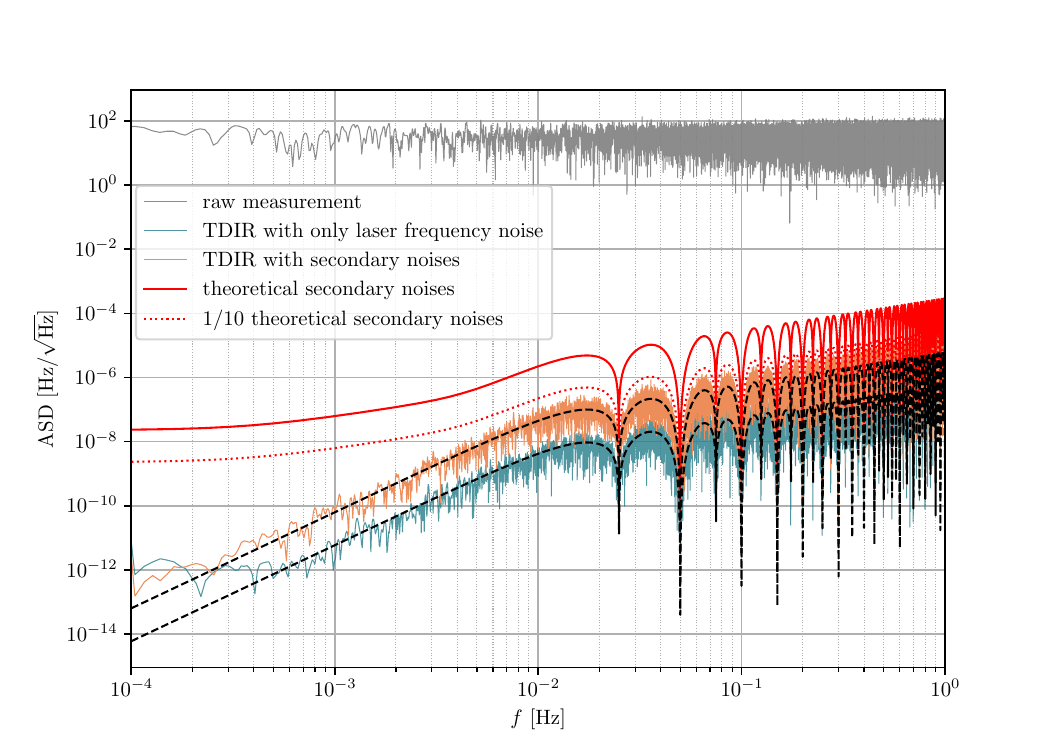} 
    \caption{Left panel:  The distributions of  $\Delta R$ calculated from the MCMC  samples of  Simulation I, II. $\Delta R_{\rm median, I} = 0.49 \ {\rm ns}$ and  $\Delta R_{\rm median, II} = 5.28 \ {\rm ns}$ are marked with vertical dashed lines. Right panel: 
    The LFN reduction effect achieved by the median values of parameters. 
    Results shown here are  also based on a 
    LFN-only simulation, and  the ASD in grey  indicates  the original readout of  science interferometer 12.  The  black dashed lines are the  theoretical ASDs of residual LFN calculated with $\Delta R_{\rm median, I/II}$. }
    \label{fig:result_stat}
\end{figure*}


The posterior distributions obtained from the two simulations are shown in FIG.~\ref{fig:posterior}. Between the two overlapping  corner plots, results of Simulation I and II  are shown in green and orange, respectively. The inner and outer contours represent the 1 $\sigma$ and 2 $\sigma$ confidence intervals of the parameters, and the ``true values'' are marked with the red lines. In the 1D distribution plots on the diagonal, the vertical dashed lines indicate the median values and 1 $\sigma$ ranges   of  corresponding parameters. 
Unsurprisingly, the posterior distribution based on  Simulation I is  unbiased, since in the absence of the random disturbance caused by  secondary noises,  parameters that lead to  the maximum ${\rm ln}\mathcal{L}$) (i.e. minimum noise) should coincide with the true values. 
As for  Simulation II, the median values of parameters are  shifted relative to the true values, but generally  still within the 3 $\sigma$  ranges.
Through the comparison between  two simulations, we can justify that these shifts are not caused by  improper  parametrization, but rather due to the impact of the  specific  realization of secondary noises. 
The strong degeneracy among the polynomial coefficients describing each arm (e.g.  $\{A_{12}, B_{12}, C_{12}\}$) is a reasonable outcome, as the delays are represented by their combinations. While, no evident correlation between the arms is observed, coincident with the theoretical analysis using  a single  $X_2$ channel. 

How to evaluate the accuracy of ranging based on the posterior distribution  can be a perplexing issue. This is because the formulas and threshold presented in Sec.~\ref{subsec:requirement}  are only suitable for  constant biases, whereas we model the delays as functions of  time. The deviations  of $C_{ij}$ relative to the true values can be up to 100 ns, which seems to suggest a pessimistic conclusion. However, these parameters only represent the initial delays and does not encompass how the fitting functions performs over the whole duration of a day. Therefore, in order to equivalent the time-varying delays to  constant biases, hence gaining an intuitive understanding of the results, we define the average  ranging error based on the  effect of LFN suppression  following  Eq.~(\ref{eq:laser_residual}) and Eq.~(\ref{eq:avg_bias}):
\begin{equation}
    \Delta R \equiv \sqrt{\frac{1}{3N}\sum_{X \in \{X_2, Y_2, Z_2\}} \sum_i^N \frac{{P}^{\rm L, MCMC}_X(f_k)}{P^{\rm L}_{X}(f_k; \Delta R =1)}}.
\end{equation}
For each MCMC sample, we calculate the corresponding delays with Eq.~(\ref{eq:parametrization}). The delays are then applied to a LFN-only simulated data (not necessarily Simulation I)   to combine TDI channels $\{X_2, Y_2, Z_2\}$, and  ${P}^{\rm L, MCMC}_X(f_k)$ is the resultant PSD of channel $X$ at frequency $f_k$ . Ratio ${P}^{\rm L, MCMC}_X(f_k)/P^{\rm L}_{X}(f_k; \Delta R =1)$
 is then averaged across frequencies $10^{-3} \ {\rm Hz}<f_k<2.4\times 10^{-2} \ {\rm Hz}$ (we set this upper limit to avoid modeling errors at the characteristic frequencies) and TDI channels $\{X_2, Y_2, Z_2\}$ to construct   an average measurement of the ranging bias. We have chosen the original Michelson channels instead of the optimal ones since they offer an intuitive way to define an averaged bias $\Delta R$.

 The left panel of FIG.~\ref{fig:result_stat} shows the distributions of $\Delta R$ calculated from the MCMC  samples of  Simulation I and  II. The  $\Delta R$s corresponding to the median values of parameters    are represented  by the vertical dashed lines and denoted as $\Delta R_{\rm median, I/II}$. 
 For Simulation I, $\Delta R_{\rm median, I} = 0.49 \ {\rm ns}$ appears at the minimum of the distribution.  While, for Simulation II, due to the disturbance cause by the secondary noises,  $\Delta R_{\rm median, II} = 5.28 \ {\rm ns}$ does not coincide with the best LFN suppression effect among the posterior samples, but rather occurs in  the lower-middle part of the distribution. Additionally, each  distribution of $\Delta R$ exhibit a dispersion of around 2 ns, which,  in a sense, can be regarded as a measure for the uncertainty of TDIR.

The LFN reduction effects achieved by the median values of parameters are shown in  the right panel of FIG.~\ref{fig:result_stat}. These 
results  are  also based on a LFN-only simulation, and  the grey ASD plot  indicates  the original readout of  science interferometer 12.  The  black dashed lines are the  theoretical ASDs of residual LFN calculated with $\Delta R_{\rm median, I/II}$. The agreement  between theory and simulation  suggests that our definition of  the average  ranging bias is reasonable. 
Judging from the threshold set by 1/10 of the secondary noises, we draw a  conclusion that TDIR remains feasible in the  presence of secondary noises. 

In all the aforementioned investigation, for the sake of generality, we have assumed that laser  locking is not implemented hence all six lasers are independent of each other. While, in practice, once a specific locking scheme is set,  the six lasers  become correlated, which would consequently affect the residual LFN after TDI, and the covariances between different  arms obtained by TDIR.
{\color{black} Taking the locking scheme N1-LA12~\cite{Olaf_PhD_thesis} as an example, for the equal-arm case, the residual LFN in the Michelson $X_2$ channel caused by ranging errors reads: 
\begin{eqnarray}
    P^{\rm L}_{X_2}(f; \Delta R_{ij}) &=& 16\sin^2(2u)\sin^2(u) \omega^2 P_{\rm L}(f)  \nonumber \\ 
    &\times& \left(\Delta R_{13} + \Delta R_{31} - \Delta R_{12} - \Delta R_{21}\right)^2. \label{eq:laser_residual_lock}
\end{eqnarray}
And the PSD of the secondary noises remains the same as Eq.~(\ref{eq:TDIX2ndnoise}). With Eq.~(\ref{eq:Fisher}), one can easily verify that the Fisher information matrix is singular,  indicating  that there is degeneracy among  the parameters. 
In Appendix~\ref{sec:locking}, we offer a  glimpse at  the  posterior distributions of delay parameters in the ``locked lasers'' case, compared to the ``no locking'' case. The figure shows that 
the combination of different TDI channels can mitigate the degeneracy to a certain extent, but the correlation between different arms still exists.}  Furthermore, 
the differences introduced by various locking schemes, as well as the selection of ``optimal'' scheme in the sense of TDIR  remains to be comprehensively explored in future research.

\section{Conclusions}\label{sec:conclusions}
In this paper, we  explored the feasibility and performance of  TDIR as a ranging approach independent  of the PRN signal, considering  the realistic  scenarios including  LFN, secondary noises, OB noise, clock noise, orbital motion, and clock desynchronization. MCMC sampling based on simulated data   yielded a 5.28 ns  ranging bias together with  a 2 ns  dispersion,  lower than the threshold which is set according to the effect of LFN suppression. 
 
Our research has shown promising prospects for the application of TDIR, which may serve as a cross-validation or backup for the traditional PRN code-based method. This approach is free of the imperfections of PRNR, and hence the sophisticated PRNR processing steps. Additionally, the whole TDI data processing flow based on our TDIR algorithm  has more relaxed  requirement on the accuracy of  clock synchronization. 

Meanwhile, 
there are at least {\color{black}four} issues that remain to be addressed in the future. 
Firstly, 
in our research, running the  two-phase MCMC sampling   took  $\sim \mathcal{O}(5)$ days on 
a personal laptop,  highlighting the imperative to enhance both the  computational power and the efficiency of  algorithm. 
{\color{black}For example, using an optimization algorithm to perform a maximum likelihood estimation (MLE)
may be a more efficient approach.
After all, in practical applications, we are more concerned with the optimal values of ranges rather than their posterior distributions.}

Secondly, since TDIR takes the interferometric data as the input, the compositions of data have  significant impacts on the outcome. 
{\color{black}Although we have tried to avoid  the encounter with GW signals by only using the high-frequency data, the possibility that there might still be signals can not be ruled out.}
In the presence of GW signals, it is necessary to incorporate TDIR into the global fit pipeline, and the effect of joint estimation with GW source parameters  necessitates further investigation. 

{\color{black}Thirdly, in the calculation of likelihood function, we have assumed that the noise models $P^{\rm 2nd}_X(f)$, as functions of frequency and armlengths, are known a prior, indicating that the secondary noises have been well calibrated. While, in practice, due to the variations of instruments and space environment, the noise models may evolve over time, thus TDIR may confront the situation of inaccurate or unknown noise models. 
In face of such situation, an  iterative estimation may be a possible solution. To be specific, the 1st step is a MLE with $P^{\rm 2nd}_X(f) \equiv 1$, then we construct TDI combinations using the resulting TDIR, and subsequently evaluate the PSDs of these TDI combinations. These PSDs will serve as  $P^{\rm 2nd}_X(f)$ and be input to the 2nd TDIR estimation step.  As such, the iterative process of refining the estimates of TDIR and PSD will continue incrementally until convergence.}

At last, the majority of our research assumed six independent lasers, and  a simple illustrative example was  provided for the case with laser locking. 
More detailed  discussion regarding the impacts and choice of  locking schemes also awaits  future research.

\begin{acknowledgments}
This study is supported by the National Key Research and Development Program of China (Grant No. 2021YFC2201903, Grant No. 2021YFC2201901 and Grant No. 2020YFC2200100).
\end{acknowledgments}

\bibliography{apssamp}

\appendix
\begin{widetext}

\section{Noise models}\label{sec:appendix}
All instrumental noises are assumed to be Gaussian and stationary, and the amplitudes and shapes of PSDs are chosen according to the   current  requirements for the Taiji mission. 
\begin{itemize}
    \item LFN:
        \begin{equation}
            P_{\rm L}(f) = A_{\rm L}^2,
        \end{equation}
        with $A_{\rm L} = 30 \ {\rm Hz/\sqrt{Hz}}$.
    \item OB noise:
        \begin{equation}
            P_{\rm OB}(f) = A_{\rm OB}^2,
        \end{equation}
        with $A_{\rm OB} =  5.65\times 10^{-18} \ {\rm s/\sqrt{Hz}}$, equivalent to a nanometer level displacement jitter, or a  10 mrad level phase disturbance.
    \item clock noise:
        \begin{equation}
            P_{\rm clock}(f) = A_{\rm clock}^2 \left(\frac{{\rm 1 \ mHz}}{f}\right)^3,
        \end{equation}
        where $A_{\rm clock} = 3.16 \times 10^{-10} \ {\rm s/\sqrt{Hz}}$.
    \item secondary noises:

        \begin{eqnarray}\label{eq:2ndnoise}
            P_{\rm OMS}(f) &=& A_{\rm OMS}^2 \left[1 + \left(\frac{2 \ {\rm mHz}}{f}\right)^4\right],  \\
            P_{\rm ACC}(f) &=& A_{\rm ACC}^2   \left[1 + \left(\frac{0.4 \ {\rm mHz}}{f}\right)^2\right] \left[1 + \left(\frac{f}{8 \ {\rm mHz}}\right)^4\right]. 
        \end{eqnarray}
        To achieve the target  sensitivity of Taiji, the required  amplitudes of optical measurement noise and test-mass acceleration noise  are  $A_{\rm OMS}=8 \ {\rm pm/\sqrt{Hz}}$ and $A_{\rm ACC} = 3 \ {\rm fm / s^2 / \sqrt{Hz}}$, respectively. Note that $P_{\rm OMS}(f)$ and $P_{\rm ACC}(f)$ should be multiplied by $(2\pi f  \nu_0 / c)^2$ and $(\nu_0 / (2\pi f c))^2$ to be converted to  frequency unit before inserting into Eq.~(\ref{eq:TDIX2ndnoise}).
\end{itemize}

The PSDs of secondary noises in the Michelson and optimal TDI channels for the general unequal-arm scenario read
\begin{eqnarray}
    P^{\rm 2nd}_{X_2}(f) &=& 32 \sin^2 \left(\frac{u_{12131}}{2}\right)  \left\{P_{\rm OMS}(f)\left[\sin^2 \left(\frac{u_{121}}{2}\right)+\sin^2 \left(\frac{u_{131}}{2}\right)\right] \right. \nonumber \\
    &&+ 2 P_{\rm ACC}(f) \left[\sin^2 \left(\frac{u_{121}}{2}\right)\left(\cos^2 \left(\frac{u_{131}}{2}\right)+1\right) \right.  + \left.\left.\sin^2 \left(\frac{u_{131}}{2}\right)\left(\cos^2 \left(\frac{u_{121}}{2}\right)+1\right)\right]\right\}, \label{eq:X2_unequal}\\
    P_{X_2Y_2^*}^{\rm 2nd}(f) &=& -(1-\Delta_{131})(1-\Delta_{12131})(1-\Delta_{232}^*)(1-\Delta_{23212}^*)  (\Delta_{21}^*+\Delta_{12})\left[P_{\rm OMS}(f) + 4P_{\rm ACC}(f)\right], \label{eq:X2Y2_unequal}\\ 
    P_{A_2}^{\rm 2nd}(f) &=& \frac{1}{2}\left[P^{\rm 2nd}_{X_2}(f) + P^{\rm 2nd}_{Z_2}(f)\right] - {\rm Re}\left[P_{Z_2X_2^*}^{\rm 2nd}(f)\right],  \label{eq:A2_secondary}\\
    P_{E_2}^{\rm 2nd}(f) &=& \frac{1}{6}\left\{P^{\rm 2nd}_{X_2}(f) + 4P^{\rm 2nd}_{Y_2}(f) + P^{\rm 2nd}_{Z_2}(f) + {\rm Re}\left[2 P_{Z_2X_2^*}^{\rm 2nd}(f) -4 P_{X_2Y_2^*}^{\rm 2nd}(f) - 4P_{Y_2Z_2^*}^{\rm 2nd}(f)\right] \right\},  \\
    P_{T_2}^{\rm 2nd}(f) &=& \frac{1}{3}\left\{P^{\rm 2nd}_{X_2}(f) + P^{\rm 2nd}_{Y_2}(f) + P^{\rm 2nd}_{Z_2}(f) + 2{\rm Re}\left[P_{Z_2X_2^*}^{\rm 2nd}(f) + P_{X_2Y_2^*}^{\rm 2nd}(f) + P_{Y_2Z_2^*}^{\rm 2nd}(f)\right] \right\}, \label{eq:T2_secondary}
\end{eqnarray}
where $u_{i_1...i_n} \equiv 2 \pi f (d_{i_1i_2} + ... + d_{i_{n-1}i_n})$, $d_{i_1i2}, ... , d_{i_{n-1}i_n}$ being the LTTs, and  $\Delta_{i_1...i_n} \equiv \exp(-i u_{i_1...i_n})$. The expressions for   $Y_2$, $Z_2$, $Y_2Z_2^*$ and $Z_2X_2^*$ can be obtained from Eq.~(\ref{eq:X2_unequal}) and Eq.~(\ref{eq:X2Y2_unequal}) via cyclical permutation of the indices : $1 \rightarrow 2$, $2 \rightarrow 3$, $3\rightarrow 1$.

We also present the PSDs of residual LFN in the optimal channels due to ranging errors, as a complement to the $X_2$ channel discussed in the main text:
\begin{eqnarray}
    P^{\rm L}_{A_2}(f; \Delta R_{ij}) &=& 8\sin^2(2u)\sin^2(u) \omega^2 P_{\rm L}(f) 
     \left[\Delta R_{12}^2 + \Delta R_{21}^2 + \Delta R_{23}^2 + \Delta R_{32}^2 + 4\cos^2\frac{u}{2}(\Delta R_{31}^2 + \Delta R_{13}^2)\right], \\
    P^{\rm L}_{E_2}(f; \Delta R_{ij}) &=& \frac{8}{3}\sin^2(2u)\sin^2(u) \omega^2 P_{\rm L}(f) \nonumber \\
    && \times
     \left[4\sin^2\frac{u}{2}(\Delta R_{13}^2 + \Delta R_{31}^2) + (5+4\cos u)(\Delta R_{23}^2 + \Delta R_{32}^2 + \Delta R_{12}^2 + \Delta R_{21}^2)\right], \\
    P^{\rm L}_{T_2}(f; \Delta R_{ij}) &=& \frac{64}{3}\sin^2(2u)\sin^2(u) \omega^2 P_{\rm L}(f) 
      \sin^2\frac{u}{2}\left(\Delta R_{12}^2 + \Delta R_{21}^2 + \Delta R_{13}^2 + \Delta R_{31}^2 + \Delta R_{23}^2 + \Delta R_{32}^2\right).
\end{eqnarray}
These expressions are derived under the equal-arm approximation and  only used in order-of-magnitude analysis.

\section{An illustrative example for locked lasers}\label{sec:locking}

\begin{figure*}
\includegraphics[width=\textwidth]{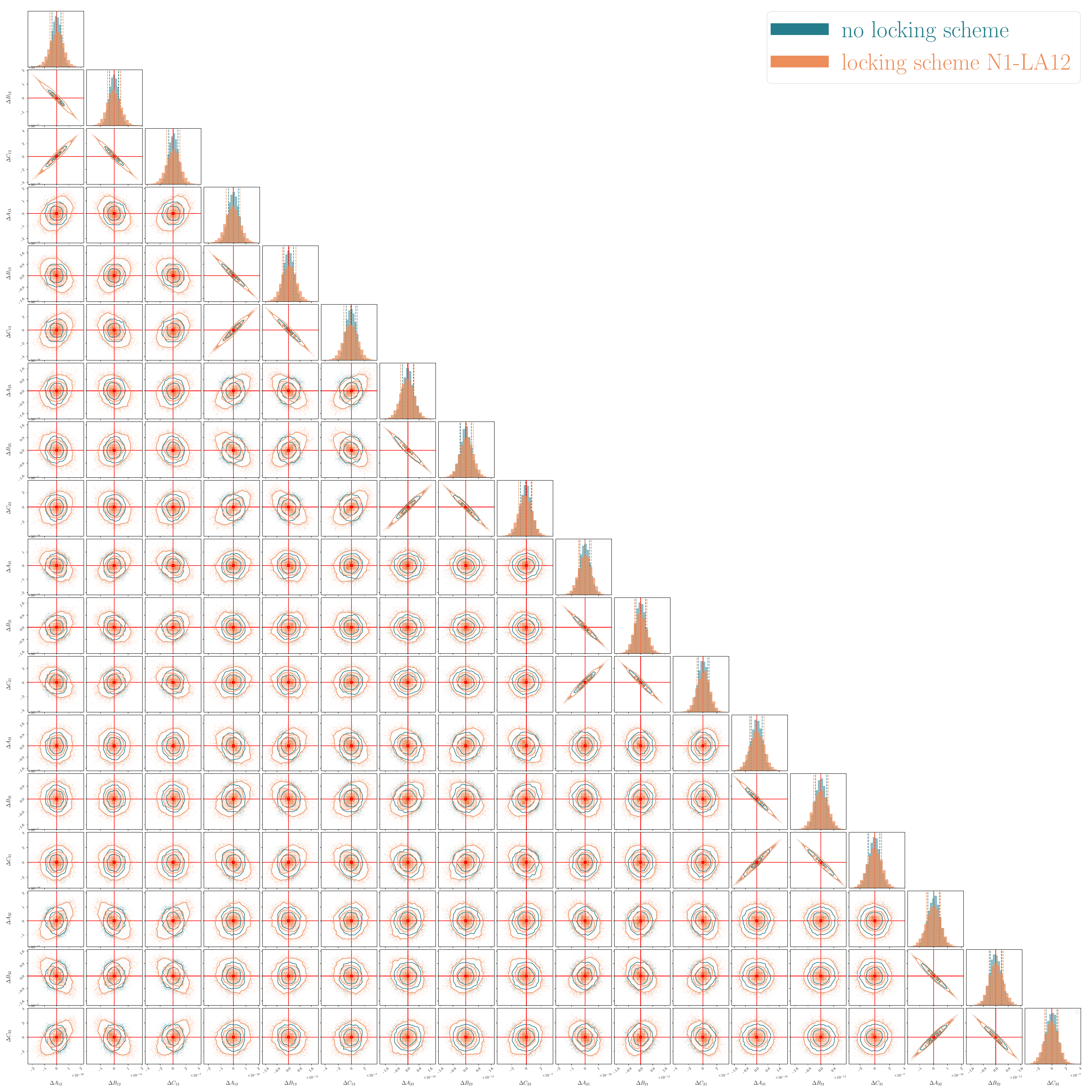}
\caption{\label{fig:locking} Similar to FIG.~\ref{fig:posterior}, but the   green and orange corner plots represent the posteriors of the ``no locking'' and ``N1-LA12 locking scheme'' cases, respectively. Both results are obtained based on  LFN-only simulations.}
\end{figure*}

\end{widetext}

\end{document}